%% file: main.tex
\documentclass[format=acmsmall]{acmart}

\AtBeginDocument{%
  \providecommand\BibTeX{{%
    \normalfont B\kern-0.5em{\scshape i\kern-0.25em b}\kern-0.8em\TeX}}}

\usepackage{xac}

\setcopyright{acmcopyright}
\copyrightyear{2022}
\acmYear{2022}
\acmDOI{10.1145/nnnnnnn.nnnnnnn}

\begin{document}

\title{Improving Workflow Integration with xPath: Design and Evaluation of a Human-AI Diagnosis System in Pathology}
\newcommand{\xp}[0]{{\sc xPath}}



\author{Hongyan Gu}
\email{ghy@ucla.edu}
\authornote{Corresponding authors.}
\orcid{0000-0001-8962-9152}
\author{Yuan Liang}
\email{liangyuandg@ucla.edu}
\orcid{0000-0002-1861-4479}
\author{Yifan Xu}
\orcid{0000-0001-8161-7317}
\email{yifanxu@ucla.edu}
\affiliation{%
  \streetaddress{580 Portola Plaza, Room 1538}
  \institution{University of California, Los Angeles}
  \city{Los Angeles}
  \state{California}
  \postcode{90095}
  \country{USA}}

\author{Christopher Kazu Williams}
\email{ckwilliams@mednet.ucla.edu}
\orcid{0000-0003-3253-7630}
\author{Shino Magaki}
\email{smagaki@mednet.ucla.edu}
\orcid{0000-0003-0433-5759}
\author{Negar Khanlou}
\orcid{0000-0002-7308-9363}
\email{nkhanlou@mednet.ucla.edu}
\author{Harry Vinters}
\orcid{0000-0002-2965-1240}
\email{hvinters@mednet.ucla.edu}
\author{Zesheng Chen}
\orcid{0000-0002-5413-1415}
\email{zesheng.chen.med1@ssss.gouv.qc.ca}
\affiliation{%
  \institution{UCLA David Geffen School of Medicine}
  \city{Los Angeles}
  \state{California}
  \country{USA}
  \postcode{90095}
}

\author{Shuo Ni}
\orcid{0000-0002-4125-2373}
\email{shuoni@usc.edu}
\affiliation{%
  \institution{University of California, Los Angeles and University of Southern California}
  \city{Los Angeles}
  \state{California}
  \postcode{90095}
  \country{USA}
}

\author{Chunxu Yang}
\email{chunxuyang@ucla.edu}
\orcid{0000-0002-9684-7534}
\affiliation{%
  \institution{University of California, Los Angeles and Peking University}
  \city{Los Angeles}
  \state{California}
  \postcode{90095}
  \country{USA}
}

\author{Wenzhong Yan}
\orcid{0000-0002-3711-0807}
\email{wzyan24@ucla.edu}
\affiliation{%
  \institution{University of California, Los Angeles}
  \city{Los Angeles}
  \state{California}
  \postcode{90095}
  \country{USA}
}

\author{Xinhai Robert Zhang}
\orcid{0000-0001-7524-2547}
\email{Xinhai.R.Zhang@uth.tmc.edu}
\affiliation{%
  \institution{University of Texas Health Science Center at Houston}
  \city{Houston}
  \state{Texas}
  \postcode{77030}
  \country{USA}}

\author{Yang Li}
\orcid{0000-0001-7808-5524}
\email{yangli@acm.org}
\affiliation{%
  \institution{Google Research}
  \city{Mountain View}
  \state{California}
  \postcode{94043}
  \country{USA}}

\author{Mohammad Haeri}
\authornotemark[1]
\email{mhaeri@kumc.edu}
\orcid{0000-0001-6055-9779}
\affiliation{%
  \institution{University of Kansas Medical Center}
  \city{Kansas City}
  \state{Kansas}
  \postcode{66160}
  \country{USA}}

\author{Xiang `Anthony' Chen}
\email{xac@ucla.edu}
\orcid{0000-0002-8527-1744}
\affiliation{%
  \institution{University of California, Los Angeles}
  \city{Los Angeles}
  \state{California}
  \postcode{90095}
  \country{USA}}

\renewcommand{\shortauthors}{Gu, et al.}

\begin{abstract}
\input{00_abstract}
\end{abstract}



\maketitle

\input{01_introduction}
\input{02_related_work}
\input{03_background}
\input{04_framework}
\input{05_design}
\input{06_implementation}
\input{07_evaluation}
\input{08_finding}
\input{10_discussion}
\input{11_conclusion}
\bibliographystyle{ACM-Reference-Format}
\bibliography{99_reference,99_added_reference}

\appendix
\input{13_appendix}

\end{document}

%% file: 00_abstract.tex
Recent developments in AI have provided assisting tools to support pathologists' diagnoses. However, it remains challenging to incorporate such tools into pathologists' practice; one main concern is AI's insufficient workflow integration with medical decisions. We observed pathologists' examination and discovered that the main hindering factor to integrate AI is its incompatibility with pathologists' workflow. To bridge the gap between pathologists and AI, we developed a human-AI collaborative diagnosis tool --- \xp~--- that shares a similar examination process to that of pathologists, which can improve AI's integration into their routine examination. The viability of \xp~is confirmed by a technical evaluation and work sessions with twelve medical professionals in pathology. This work identifies and addresses the challenge of incorporating AI models into pathology, which can offer first-hand knowledge about how HCI researchers can work with medical professionals side-by-side to bring technological advances to medical tasks towards practical applications.

%% file: 01_introduction.tex
\section{Introduction}

The past decade has experienced rapid development in digital pathology, which transforms physical glass slides into high-resolution digital whole slide images (WSIs) \cite{pantanowitz2011review}. This transformation lays the foundation for assisting diagnoses with machine intelligence \cite{amgad2019structured, bulten2021artificial, han2021multi}, and might improve patient management ultimately \cite{bera2019artificial}. To date, AI (Artificial Intelligence) has been proposed for a broad spectrum of potential applications of pathology \cite{wang2016deep, huang2018improving, rakhlin2018deep, strom2020artificial, wang2019artificial}, with some achieving performance on par with human beings in labs \cite{bejnordi2017diagnostic, zhang2019pathologist}. Furthermore, various AI models have been adopted into tools to support pathologists' tasks, targeting automating parts of pathologists' workflow to reduce their examination burdens \cite{cai2019hello, lindvall2021rapid, dov2021hybrid}.
However, it is still challenging to convince pathologists to transform from manual diagnosis to AI-based methods in practice. We believe this is caused by the dichotomy between AI and medical communities --- while the existing medical AI research focuses on improving performance, there is a lack of understanding of how doctors could benefit from AI and effectively use it for diagnosis \cite{yang2016investigating, maniatopoulos2015moving, tizhoosh2018artificial, khairat2018reasons}.

This onerous issue --- the need to integrate AI-based tools into the medical workflow --- has recently gained extensive attention in the HCI community. Empirical studies have interviewed medical professionals about their attitude toward using AI in practice, and suggest that medical systems should ``state explicitly on how AI benefits users'' \cite{cai2019hello} and ``connect to existing clinical processes'' \cite{jacob2021design}; it also indicates ``unique difficulties'' in converting human-AI interaction guidelines to tool support \cite{yang2016investigating}. To this end, previous literature has explored the designs and influence of human-AI collaborative workflows for medical professionals \cite{beede2020human, fogliato2022goes, lee2021human, wang2021brilliant}. For pathology, numerous works have revealed the potential of human-AI collaborative systems to support doctors' exploration of one or more pathological patterns \cite{cai2019human, lindvall2021rapid, corvo2017pathova}. Extending the success of previous works, this work focuses on pathologists' more complicated diagnosis tasks, and studies how interfaces should be appropriately designed between pathologists and AI to address the workflow integration challenge, given the AI's incompatibility with existing pathologists' diagnosis workflow.

To reveal how AI-aided systems should be designed, we first conducted a formative study with four experienced pathologists (average experience $\mu=21.25$ years) and summarized the main findings into the following design challenges:

\begin{enumerate}
    \item \textbf{Comprehensiveness}. Previous pathology decision support systems assist perspectives of pathologists' tasks, such as searching for one/more pathological patterns \cite{lindvall2021rapid}, or assisting adjudications on areas of interest \cite{hegde2019similar, cai2019human}. However, it is still challenging for the current systems to support diagnoses with multiple criteria from multiple pathological tests. This requires AI-aided pathology systems to comprehensively incorporate multiple criteria through a tight collaboration with pathologists;
    
    \item \textbf{Explainability}.  Previous eXplainable AI (XAI) research interprets AI predictions using explainable elements, such as attention maps \cite{zhang2019pathologist}, concept attributions \cite{cai2019human}, and confidence scores \cite{evans2022explainability}. However, it is still unclear how to effectively employ these components in pathologists' diagnosis, a time-sensitive but high-stakes process. In practice, pathologists expect to trace an AI-generated diagnosis to abundant evidence that explains such a decision;
    
    \item \textbf{Integrability}. Because of the complexity and the uncertainty of AI's output \cite{yang2020re}, it is challenging to present AI's comprehensive findings with explanations to match the diagnosis workflow of pathologists without incurring extra cognitive burdens, given the importance difference in each finding to the diagnosis according to the medical guidelines \cite{louis20072007}.
\end{enumerate}

\begin{figure}
    \centering
    \includegraphics[width=1.0\linewidth]{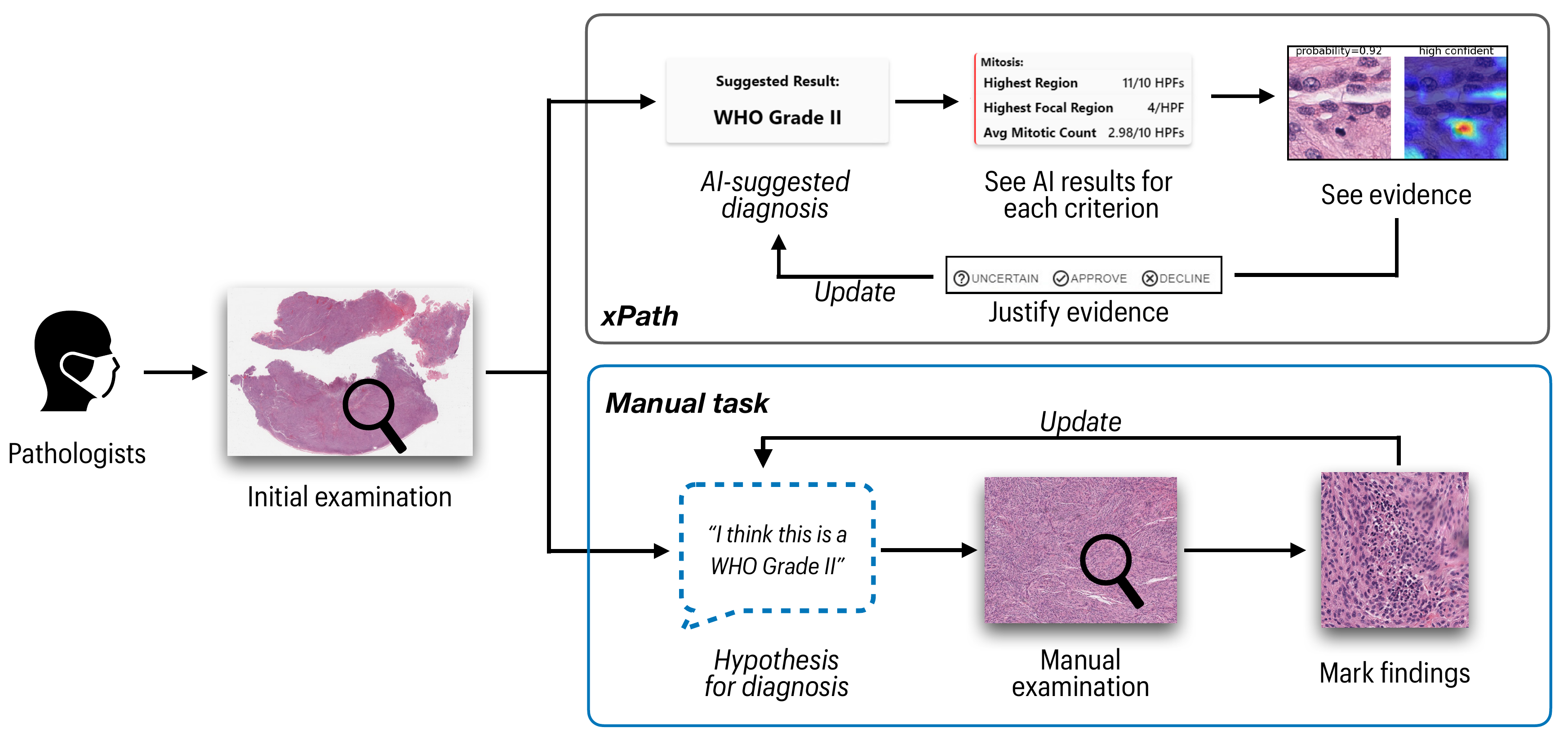}
    \caption{Workflow of \xp~(up): pathologists first see the AI-suggested diagnosis, then examine its results and evidence accordingly in an explainable manner, and examine the evidence to update the suggested diagnosis. This workflow follows a similar manual examination process of pathologists (down), which can improve AI's integration into pathologists' routine diagnoses.}
    \label{fig:workflow}
\end{figure}

\begin{figure}[t]
    \centering
    \includegraphics[width=\linewidth]{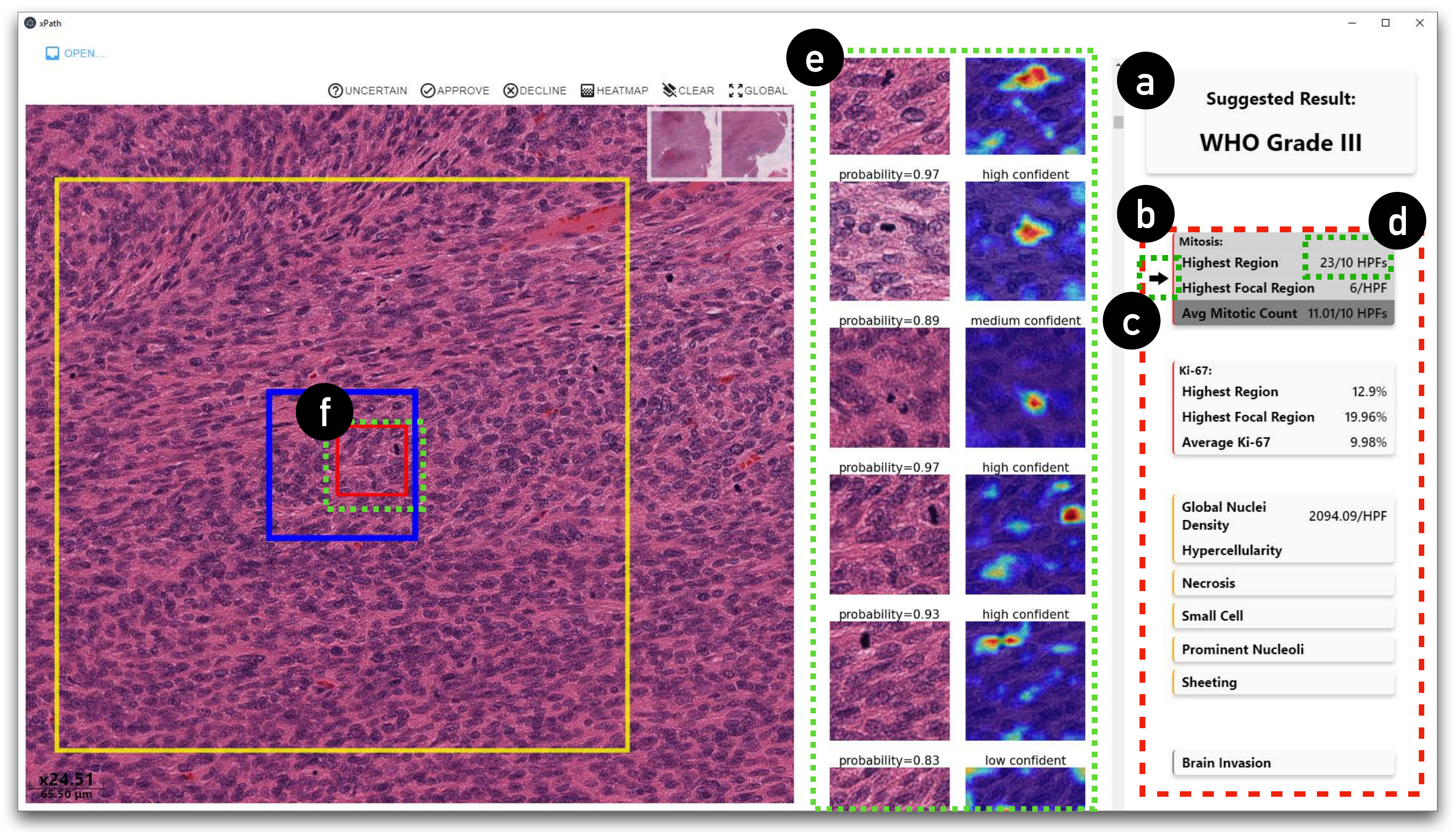}
    \caption{\xp's interface design, illustrating the (a) suggested pathology diagnosis (\ie WHO Grade 3) with two key design ingredients of (b) joint-analyses of multiple criteria, where \xp~offers comprehensive AI analysis of multiple critical pathology criteria for a diagnosis; explanation by hierarchically traceable evidence, explaining high-level suggested diagnosis to low-level AI-reporting on each pathological feature, including (c) an arrow that points to the deterministic criterion for the suggested diagnosis, (d) a quantified score for the criterion, (e) a list of evidence that contributes the quantified score, and (f) each piece of evidence registered to the whole slide image to support pathologists' examination with contextual information.}
    \label{overview2}
\end{figure}

Building upon the design challenges from the formative study, we propose \xp~--- a comprehensive and explainable human-AI collaborative diagnosis tool that can assist pathologists' examinations integrated into their practice. Specifically, \xp~can enhance pathologists' workflow integration with AI-based diagnosis from three aspects: \one it reports multiple AI-computed pathology criteria, which are critical for diagnosis according to medical guidelines; \two it presents traceable evidence for each AI report, making it accountable and explainable; \three it allows pathologists to perform diagnoses in a similar workflow to their routine practice (as shown in Figure \ref{fig:workflow}).

We realize \xp~with two design ingredients: \textbf{joint-analyses of multiple criteria} and \textbf{explanation by hierarchically traceable evidence}. First, the joint-analyses of multiple criteria present AI's findings based on multiple juxtaposed criteria from two pathology tests (Figure \ref{overview2}b), which are combined to produce a suggested diagnosis (Figure \ref{overview2}a) based on rules derived from the existing medical guideline \cite{louis20072007}. Such a design addresses the comprehensiveness challenge, where pathologists are supported by AI-results of multiple criteria. Second, the design of hierarchically traceable evidence establishes a chain of accountable evidence for the diagnosis, explaining multiple levels of AI results, from high-level suggested diagnosis, to mid-level AI's reporting on each pathological pattern, and further to each piece of evidence: a user can trace the suggested diagnosis (Figure \ref{overview2}a) with a quantified score for the criterion (Figure \ref{overview2}d), to a list of evidence that contributes to the quantified score (Figure \ref{overview2}e), and further to examine each evidence with contextual information by registering it to the whole slide image (Figure \ref{overview2}f). Such a design addresses the explainability challenge by making the provenance of a criterion traceable and transparent. With the two designs, pathologists are freed from examining the pathology data with manual exploration of the high-resolution whole slide image, but building upon their diagnosis based on their seeing, understanding, and verifying AI results. Such a workflow with AI is also similar (and thus can be integrable) to pathologists' in practice (see Figure \ref{fig:workflow}).

As for the validation of \xp, we hosted work sessions with twelve medical professionals in pathology\footnote{, which includes two attendings, two fellows, seven senior residents, and one junior resident.} across three medical centers in the United States. We used data from a local medical center and asked our participants to diagnose with the same examination protocol as they had done in practice. We used working systems of \xp~and an off-the-shelf whole slide image viewer as the baseline. Our observations found that, with less than one hour's learning, participants could effectively utilize \xp~to perform diagnosis. Specifically, they could use \xp's multi-criteria analysis by prioritizing one criterion and referring to others on demand. Furthermore, \xp's design of hierarchical explainable evidence enables participants to navigate between high-level AI results and low-level pathological details. A post-study questionnaire shows that, compared to the baseline system, participants reported \xp~more integrable with their existing workflow ($p$=0.006, Wilcoxon rank-sum test, same below): they were more likely to use \xp~in the future ($p$=0.002), and gave more overall preference on \xp~(\ie 9/12 participants ``totally prefer'' using \xp~than the baseline interface, and 3/12 ``much more prefer'' using \xp).

Benefiting from \xp's better workflow integration, participants reported \xp~required less effort ($p$=0.002), and was more effective in reducing the workload ($p$=0.002) in performing diagnosis. Meanwhile, participants could make more accurate diagnosis decisions with \xp, where they gave 17/20 cases correct diagnosis using \xp, compared to 7/12 correct with the baseline interface.

\subsection{Contributions}

Our main contribution is two-fold: \one throughout interviews with experienced pathologists, we identified their challenges in practice, and summarized that comprehensiveness, explainability, and integrability are the three key components for incorporating AI models into pathologists' workflow; \two based on the empirical findings, we proposed a human-AI diagnosis tool --- \xp~--- that facilitates pathologists' routine examinations collaboratively, validated by a study that evaluates pathology professionals' diagnoses compared with a baseline system. Our study and findings shed light on how HCI researchers can design integrable AI-assisted systems to bring advancements to doctors' workflow.

%% file: 02_related_work.tex
\section{Related Work}
In this section, we review the related work of \xp~ from three areas: \one AI algorithms for processing pathology images, \two enhancing AI's workflow integration for medical applications, and \three human-AI collaborative tools for pathologists.

\subsection{Processing Pathology Images with Data-Driven AI}
With the recent development of digital pathology techniques, a considerable amount of datasets have grown around the theme of marking pathological patterns from digital pathology slides. Current datasets are primarily based on H\&E (\ie Hematoxylin and Eosin, a type of pathology staining) slides, the most commonly used stained slides for providing a detailed view of the tissue. To date, these datasets cover a broad range of pathology practices, from conducting high-level diagnostic tasks, such as identifying breast cancer metastasis \cite{litjens20181399}, to seeing low-level pathological patterns, such as mitoses \cite{veta2019predicting, roux2013mitosis}.

Such an increase in data availability in digital pathology has triggered a recent surge of data-driven techniques in a broad range of applications, such as screening negative biopsies \cite{dov2021hybrid}, carcinoma detection \cite{araujo2017classification, bejnordi2017diagnostic, bardou2018classification}, quantification of pathological features \cite{cirecsan2013mitosis, veta2015assessment, gu2022detecting}, and tumor grading \cite{ertosun2015automated, arvaniti2018automated}. It is noteworthy that some previous AI models have achieved performance on par with human beings in lab studies. For example, Zhang \etal~combined multiple neural networks, including a Convolution Neural Network (CNN), a fully connected neural network, and a Recurrent Neural Network (RNN), to diagnose urothelial carcinoma, which achieves matching diagnosis performance compared to a group of pathologists \cite{zhang2019pathologist}.

Besides H\&E slides, AI algorithms have also been devised for other pathology tests that can assist decision-making, \eg Ki-67 immunohistochemistry (IHC) tests. For example, Xing \etal~trained a fully connected convolutional network to perform nucleus detection and classification from Ki-67 slides \cite{xing2019pixel}. In more recent research, Ghahremani \etal~trained a cycleGAN network with more-precise immunofluorescence data as ground truth to improve cell-level semantic segmentation for IHC tests \cite{ghahremani2021deepliif}.

Although its broad applications and promising performance, data-driven AI in pathology has caused rising ethical concerns because of the high-stakes nature of performing diagnoses \cite{chauhan2021ethics}. Multiple works ask AI to provide algorithm transparency \cite{cai2019hello} and result accountability \cite{sendak2020human, mccradden2020patient}. And several studies have included eXplainable AI (XAI) techniques to improve the transparency of data-driven AI for pathology. For example, Gehrung \etal~employed the saliency map visualization to highlight the spacial support for the model prediction, and suggest that the saliency maps a strong agreement with pathologists' labels \cite{gehrung2021triage}. To investigate pathologists' attitudes towards XAI elements, Evans \etal~further conducted a user-oriented study and found that simple visual explanations were preferred because they were closer to pathologists' visual examinations on the slides \cite{evans2022explainability}.

Going beyond providing explanations for AI predictions, other works aim to build interpretable AI for healthcare applications. For example, Choi \etal~mimicked physicians' practice of examining electronic health records and introduced an interpretable RNN model that diagnosed by detecting patients' past visits \cite{choi2016retain}. Koh \etal~trained a concept bottleneck model that can classify X-ray images with human-understandable concept values as interpretations \cite{koh2020concept}. However, it is noteworthy that interpretable AI in the pathology imaging domain is not as popular as in general AI research. We believe this is partly related to AI training: first, interpretable AI usually requires human-annotated labels (\eg concepts) for training, while pathologists' annotations are hard to acquire \cite{schaekermann2020expert}; second, it adds difficulties to training interpretable AI because its additional interpretable constraints \cite{rudin2018please}.

The progress of the AI and XAI techniques has built fundamentals of using AI to automate pathologists' tasks without losing transparency. However, the main focus of AI in the medical domain is to improve performance, while XAI research targets to explain AI findings. We argue that it is insufficient to assist pathologists' diagnoses by only optimizing the AI algorithms or simply applying XAI designs. This is because their poor integration into the medical workflow might add burdens to pathologists, which disincentivizes them to use AI systems in practice \cite{yang2016investigating}. In this work, we seek a better understanding of pathologists' expectations of AI by working closely with a group of pathologists. Based on which we further conclude three design requirements for pathology AI systems --- comprehensiveness, explainability, and integrability --- to enhance the integration of AI-aided systems in pathology.

\subsection{Enhancing AI's Workflow Integration for Medical Applications}

In the history of medical AI systems, workflow integration has been recognized as a key value for medical users. For example, 
Teach \etal~ have studied physicians' attitudes toward clinical consultation systems and offered suggestions on computer-based decision support systems, \eg ``minimizing changes to current clinical practices'' \cite{teach1981analysis}.
Middleton \etal~have reviewed research on clinical decision support systems since 1990 and pointed out that the poor integration in clinicians' workflow is becoming a barrier preventing the application of such tools \cite{middleton2016clinical}. Yang \etal~indicated that a medical AI tool should set the explicit goal of helping medical users increase the overall quality of examination, instead of insufficiently automating a part of their work \cite{yang2016investigating}.

In the general healthcare domain, literature has attempted to enhance workflow integration by improving medical users' engagement in the design process of AI systems. For example, Sendak \etal~included medical professionals in designing and implementing a deep-learning-driven sepsis monitoring system. Based on the co-designing process, they summarized takeaways to improve workflow integration, including ``respect professional discretion'' and ``create ongoing feedback loops with stakeholders'' \cite{sendak2020human}. Jacobs \etal~further concluded that medical systems should ``offer on-demand explanations'' to address the mismatch between AI predictions and the medical guidelines \cite{jacob2021design}. 

Numerous studies have explored the potential usage, issues, and influence of employing human-AI collaborative workflows in clinical settings. For example, Beede \etal~studied socio-environmental factors that influenced AI performance, nurses' workflow, and patient experience while using a deep learning system for diabetic eye disease \cite{beede2020human}. Wang \etal~revealed challenges of usability, technical limitations, and human trust that emerged from applying an AI-powered clinical diagnostic support system \cite{wang2021brilliant}.  Fogliato \etal~discovered that demonstrating AI inference at the start of radiologists' reading of X-ray images would increase doctors' agreement \cite{fogliato2022goes}. Lee \etal~reported that the human-AI collaborative system could increase therapists' agreement on the rehabilitation assessment \cite{lee2021human}.

Narrowing down to the pathology domain,  Cai \etal~highlighted pathologists' needs for information from AI, which included the AI's capabilities measured in well-defined metrics and transparency to overcome subjectivity \cite{cai2019hello}. Gu \etal~summarized six design lessons for interactive AI systems in pathology, suggesting AI systems in pathology should ``provide the actionability of the AI guidance'' and ``narrow down to small regions of a large task space'' \cite{gu2020lessons}. 

The design conclusions and guidelines open up opportunities to enhance AI's integration into pathologists' workflows. However, there are still limited working tools that support pathologists' diagnoses in the wild, consisting of examining multiple criteria from multiple pathology tests. In this work, we aim to enhance AI's workflow integration using the task of meningioma (a type of brain tumor) grading as a case study. Specifically, we propose two designs for pathology AI systems: \textbf{joint-analyses of multiple criteria} and \textbf{explanation by hierarchically traceable evidence}. We observe how pathologists interact with these two designs and summarize recurring themes, providing first-hand information for future pathology AI system designs.

\subsection{Human-AI Collaborative Tools for Pathologists}

One way to increase AI's workflow integration for medical users is by enabling them to collaborate with AI. And enabling human-AI collaboration requires ``goal understanding, preemptive task co-management and shared progress tracking'' \cite{wang2020human}.

Recent HCI research has demonstrated numerous examples of human-AI collaboration in various general tasks, such as content creation \cite{willett2018mixed, jeon2021fashionq, jeon2021fancy}, design \cite{chen2018forte, lee2019smartmanikin}, well-being \cite{yan2022emo}, and accessibility \cite{liu2022cross}. For medical tasks, various human-AI collaboration systems have shown their validity in improving doctors' agreement \cite{bulten2021artificial, fogliato2022goes, lee2021human}, mental effort \cite{cai2019human}, and accuracy \cite{bertram2022computer}. However, literature has also suggested that AI performance might be influenced by clinical factors in the wild \cite{park2019identifying, beede2020human, wang2021brilliant}. In the pathology domain, multiple works have shown that the human + AI approach could potentially increase the quality of diagnoses. For example, Wang \etal~reported that combining AI and human diagnoses improves pathologists' performance in breast cancer metastasis classification with an $\sim$85\% reduction in human error rate \cite{wang2016deep}. More recent work by Bulten \etal~has suggested that the introduction of AI assistance increases pathologists' agreement with the expert reference standard in prostate cancer grading \cite{bulten2021artificial}.

A number of existing human-AI collaboration projects on pathology have been focused on Content-Based Image Retrieval (CBIR). With a given slide (or patch) from pathologists, such tools retrieve image examples of a similar pattern to help the decision-making. For instance, Hegde \etal~proposed a reversed image searching tool to help pathologists find image patches with similar pathological features or disease states \cite{hegde2019similar}; Cai \etal~enabled pathologists to specify custom concepts that guide the retrieval of similar annotated patches of pathological patterns \cite{cai2019human}. However, the CBIR focuses on image searching: what images to search, how to use the search results, and what to conclude according to searching results. On the other hand, diagnosing/grading carcinoma in digital pathology is more complicated, requiring pathologists to detect multiple pathological features and aggregate them according to medical standards for decision-making. And \xp~is considered a tool of Computer-Aided Diagnosis (CAD) or Clinical Decision Support System (CDSS).

Existing CAD/CDSS tools can enhance the detection in digital pathology with visualization. For example, Corvo \etal~developed PathoVA, which provided AI support for breast cancer grading by visualizing three types of clues \cite{corvo2017pathova}. The system could also track pathologists' interactions and help them generate reports. Krueger \etal~ enhanced users' exploration of multi-channel fluorescence images to support cell phenotype analysis \cite{krueger2019facetto}. Specifically, the tool maintained hierarchical statistics about the number of cell-level findings to help a user keep track of analysis and interactively update the statistics with machine learning algorithms on the fly.
These tools provide a bottom-up approach to assist pathologists in making a diagnosis: pathologists are only prompted with low-level AI-generated clues (\eg highlighting tumor cells with a segmentation map); then, the diagnosis is drawn by pathologists from fusing observations with these clues.  In contrast, \xp~allows pathologists to evaluate a case with a top-down approach: they can first see an overall grading (top-level) based on joint analyses of multiple criteria, then drill down to localized areas with traceable evidence and further to low-level patterns for verification and correction. Such a design is similar to pathologists' examining the image manually, where they first develop hypotheses and interactively refine them by adding supporting evidence.

%% file: 03_background.tex
\section{Medical Background}
\label{sec:med-background}
\begin{figure}[t]
    \centering
    \includegraphics[width=1.0\linewidth]{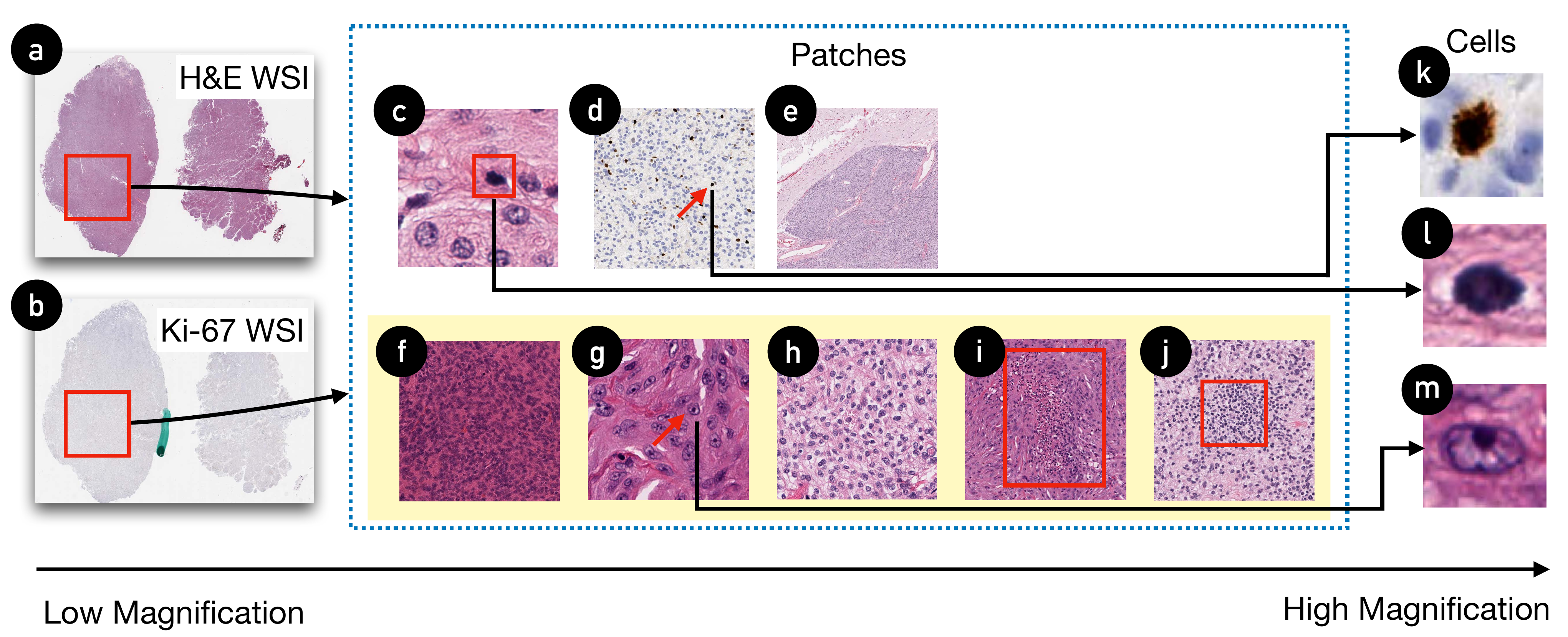}
    \caption{Examples of criteria used for the meningioma grading. (a) The resected tissues are first stained with H\&E solution. (b) An additional Ki-67 IHC test is usually used to locate mitoses. According to the WHO grading guidelines, pathologists look for (c) mitotic cells (marked in the red box) in high-power fields with the help of (d) Ki-67 stains; (e) brain invasion (invasive tumor cells in brain tissue); five pathological patterns, including (f) hypercellularity (an abnormal excess of cells), (g) prominent nucleoli (enlarged nucleoli pointed by the arrow), (h) sheeting (loss of `whirling' architecture), (i) necrosis (irreversible injury to cells marked in the red box), (j) small cells (tumor cell aggregation with high nuclear/cytoplasmic ratio marked in the red box). For some criteria, \eg mitosis (k,l) and prominent nucleoli (m), pathologists are required to zoom further into the high magnification level for examination.}
    \label{fig:meningioma}
\end{figure}

In this work, we target the task of meningioma (a type of brain tumor) grading as a case study to probe the design of human-AI collaborative tools for pathology diagnosis. The meningioma grading is selected because of its complexity --- it covers three aspects of difficulties for pathologists: \one multiple morphological and immunohistological features utilizing at least two kinds of pathology tests (\ie Hematoxylin and Eosin (H\&E) slides and Ki-67 immunohistochemistry (IHC) tests) for the grading of the tumor, \two alternate high and low magnification images to detect large structures (\ie brain invasion, see Figure \ref{fig:meningioma}e) or small events (\ie mitosis, see Figure \ref{fig:meningioma}c), and, \three examine the entire tumor (occasionally as many as 20 or more slides) for frequently rare features (\ie spontaneous necrosis, see Figure \ref{fig:meningioma}i). As such, the practice of grading meningiomas is a favorable arena for studying how human-AI collaborative systems should be designed to assist pathologists in carrying out multiplex tasks.

According to the World Health Organization (WHO) guidelines (2016), meningiomas can be graded as Grade 1, Grade 2, or Grade 3 \cite{louis20072007}. The current grading of meningioma in the new WHO guideline (2021) still recommends the same criteria for grading, although the nomenclature is slightly different. Additionally, new molecular alterations are added to determine the tumor grade \cite{louis20212021}.

The accurate grading of meningioma is vital for treatment planning: the Grade 1 tumors can be treated with either surgery or external beam radiation, while Grade 2/3 ones often need both treatments \cite{walcott2013radiation}; meanwhile, research shows that patients with Grade 3 meningiomas suffer a higher recurrence rate as well as lower survival rate in comparison to Grade 2 patients \cite{palma1997long}.

Pathologists need to search and locate multiple pathological features across various magnifications with optical microscopes or digital interfaces in order to determine the tumor grade. Specifically, they first localize the regions of interest (ROIs) in low magnification (x40), then switch to the patch level with a higher magnification (x100), and sometimes zoom further with the highest magnification (x400) to examine cellular architecture. These steps are usually repeated multiple times until pathologists have collected sufficient findings to conclude a grading and sign out the case. 

Figure \ref{fig:meningioma} briefly visualizes examples of pathological features that pathologists need to find. Pathologists' work starts with the H\&E slides (Figure \ref{fig:meningioma}a). Apart from the H\&E, Ki-67 IHC tests \cite{abry2010significance} are often used (Figure \ref{fig:meningioma}b) to provide an estimated proliferation index (Figure \ref{fig:meningioma}d,k), which is highly correlated to meningioma grading. According to the WHO guidelines\footnote{Please refer to the Appendix \ref{sec:mg-grade} for detailed descriptions of the WHO guidelines for meningioma grading.} \cite{louis20072007, louis20212021}, grading meningiomas is based on the findings of multiple microscopic or large-sized pathological features. As such, meningioma grading is challenging and high-stakes --- an overestimated study would incur unnecessary treatment on patients, and an overlooked one would cause a delay of necessary treatment.

%% file: 04_framework.tex
\section{Formative Study}
\label{sec:formative}

We conducted a formative study to reveal the system requirements for human-AI pathology diagnosis. Specifically, we recruited four experienced pathologists (average experience $\mu=21.25$ years) from a local medical center through word-of-mouth. All participants had examined meningiomas weekly. The demographic information of the participants is shown in Table \ref{tab_user}. Two out of four participants (FP3, FP4) have used digital pathology systems, and the primary software they used is Imagescope\footnote{\url{https://www.leicabiosystems.com/us/digital-pathology/manage/aperio-imagescope/}}. For familiarity with AI, one participant knows machine learning, one has passing knowledge, and two have little.

As for the process of the formative study, we started by describing the project's motivation and presented participants with a real meningioma whole slide image. Next, we asked the participants to examine the case and encouraged them to talk aloud about their examination process. We followed up with a semi-structured interview and let the participants describe the challenges in their practice and their expectations of an AI-enabled system to assist such a process. The average duration of the semi-structured interviews was about 25 minutes, and the average length of the study was about 60 minutes. Please refer to the supplementary material for the moderator's guide in the semi-structured interview.

\begin{table}
  \footnotesize
\centering
\begin{tabular}{c | c c c}
\hline
 ID & Occupation & Years of Experience & Familiarity of Meningiomas\\
\hline \hline
 FP1 & Attending/Professor & 44 & Examine Weekly\\
 FP2 & Attending/Assistant Professor & 22 & Examine Weekly\\
 FP3 & Attending & 10 & Examine Weekly\\
 FP4 & Attending & 9 & Examine Weekly\\
\hline
\end{tabular}
\caption{Demographic information of the participants in the formative study.}
\label{tab_user}
\end{table}

\subsection{Existing Challenges for Pathologists}
\label{subsec:existing}
We first transcribed the audio recordings of all interviews. One experimenter coded the transcripts and shared the recurring challenges mentioned by the participants. A second experimenter coded individually and took a pass on the first experimenter's findings. Then, a third experimenter joined to discuss with the previous two experimenters and resolved the disagreements. Resulting from the complicated the medical guideline, we discovered three challenges in the current pathology practice of meningioma grading:

\textbf{Time Consumption}. 
The small-scaled characteristics in the patterns of interest and the very high resolution of slides make the meningioma grading highly time-consuming for pathologists. A resected section from a patient's brain tissue would generate eight to twelve H\&E slides, and pathologists need to look through all those slides and integrate the information found on each slide. 
Except for the few experienced pathologists, meningioma grading can be time-consuming to go through because a single patient's case often consists of 10+ slides --- ``\textit{If you don't see obvious features of malignancy, like necrosis or mitosis, you have to search all of the slides in high power to look for mitosis, which will take a few hours}'' (FP4)
Automating portions of the slide examination process by AI can potentially reduce such time consumption, alleviate pathologists' workload, and increase the overall throughput.
  
\textbf{Subjectivity}.
There are high intra- and inter-observer variations during the grading of tumors. Pathologists summarize three factors contributing to such subjectivity:
\one a lack of precise definitions --- the WHO guidelines do not always provide a quantified description for the five pathological features of high-grade meningioma. For example, for the `prominent nucleoli' criterion, the WHO guideline does not specify how large the nucleolus should be considered as `prominent', described by FP2 --- ``\textit{... small cells, large nucleoli ... nobody has defined what that means...}'';
\two implementation of the examination process --- for example, the mitotic count for grade 2 meningioma is defined as 4 to 19 mitotic cells in 10 consecutive high-power fields (HPFs)\footnote{The size of field-of-view under x400 magnification of a optical microscope.}. However, the guideline does not specify the sampling rules of these 10 HPFs. 
As a result, different pathologists are likely to sample different areas on the slide;
\three natural variability in people, such as the level of experience, time constraint, and fatigue  \cite{croskerry2017diagnosis} --- ``\textit{One person would like to say it is mitosis, while the other person would say ‘not really', because it is not good enough.}''(FP4) For AI, the definition and implementation of guidelines can be codified into the model and visualized in the system that performs consistently to overcome people's variability. 

\textbf{Multi-Tasking}. Going beyond the time consumption and subjectivity, participants also mentioned that it was also challenging for less-experienced pathologists to ``multitask'', \ie cross-referencing amongst multiple criteria at the same time, rather than going through one after another sequentially. The ``multitasking'' operation is challenging because it requires pathologists to memorize which criterion they had found and where they were simultaneously. However, we believe such a limitation can be addressed by introducing digital systems without AI, where computers can memorize pathologists' previous annotations and interactions.

\subsection{System Requirements for xPath}
Regarding pathologists' expectations about the system, we summarized three requirements to enhance workflow integration: comprehensiveness, explainability, and integrability. Note that participants also expect the AI to be accurate and reproducible for meningioma grading --- ``\textit{If the machines cannot provide accurate material, it is not a worthwhile system ... It would be good if two different machines can give the similar quality of mitosis.}'' (FP1) However, instead of including them in the \textit{system} requirements, we believe such concerns can be addressed by the introduction of high-performance AI, which we will demonstrate in Section \ref{AI_backend}.

{\bf Comprehensiveness}.
According to the current medical guideline, the grading of meningiomas involves multiple sources of pathology tests (from H\&E and Ki-67) and criteria (\eg mitosis, necrosis, brain invasion). To incorporate \xp~ into the current practice, the system should comprehensively, systematically, and exhaustively support all these pathology tests and criteria to ensure that pathologists do not miss crucial findings.

{\bf Explainability}.
In lieu of a single grading result from a black-box AI model, the system should provide visual evidence to justify the AI's findings according to the medical definition of the criterion. This is because some criteria (only visible under high magnifications) requires examining lower-level details in order to interpret an AI's finding and further needs to be traceable to the original location in the whole slide image for a review with more contextualized information. Overall, there should be explainability both globally (how results from multiple criteria are combined to yield a grading) and locally (which includes \one what evidence leads to the computed result of each criterion, \eg where mitoses are detected that lead to the number of mitosis counts, and \two why a specific piece of evidence is captured by AI, \eg which part of the evidence convinces the AI that it contains mitoses).

{\bf Integrability} The system should allow pathologists to diagnose with AI similar to their daily routines of manual examination. Specifically, the system should first suggest a hypothesis for diagnosis and provide evidence to support it. Meanwhile, given that errors are inevitable for most existing AI models, the system should allow pathologists to refine AI's findings by retrieving detailed contextualized evidence on demand. When showing the evidence of grading, the system should not overwhelm pathologists with all evidence from a whole slide; rather, it should direct pathologists to the representative regions of interest. Finally, the system should enable pathologists to cross-check each criterion and override the results manually when they detect an error.

%% file: 05_design.tex
\section{Design of xPath}

Guided by the aforementioned system requirements, we developed \xp~ with two key designs for pathology AI systems: \one joint-analyses of multiple criteria and \two explanation by hierarchically traceable evidence. We first detail the two designs and then describe how a pathologist uses \xp~ to perform a meningioma grading task.

\begin{figure}[t]
    \centering
    \includegraphics[width=0.75\linewidth]{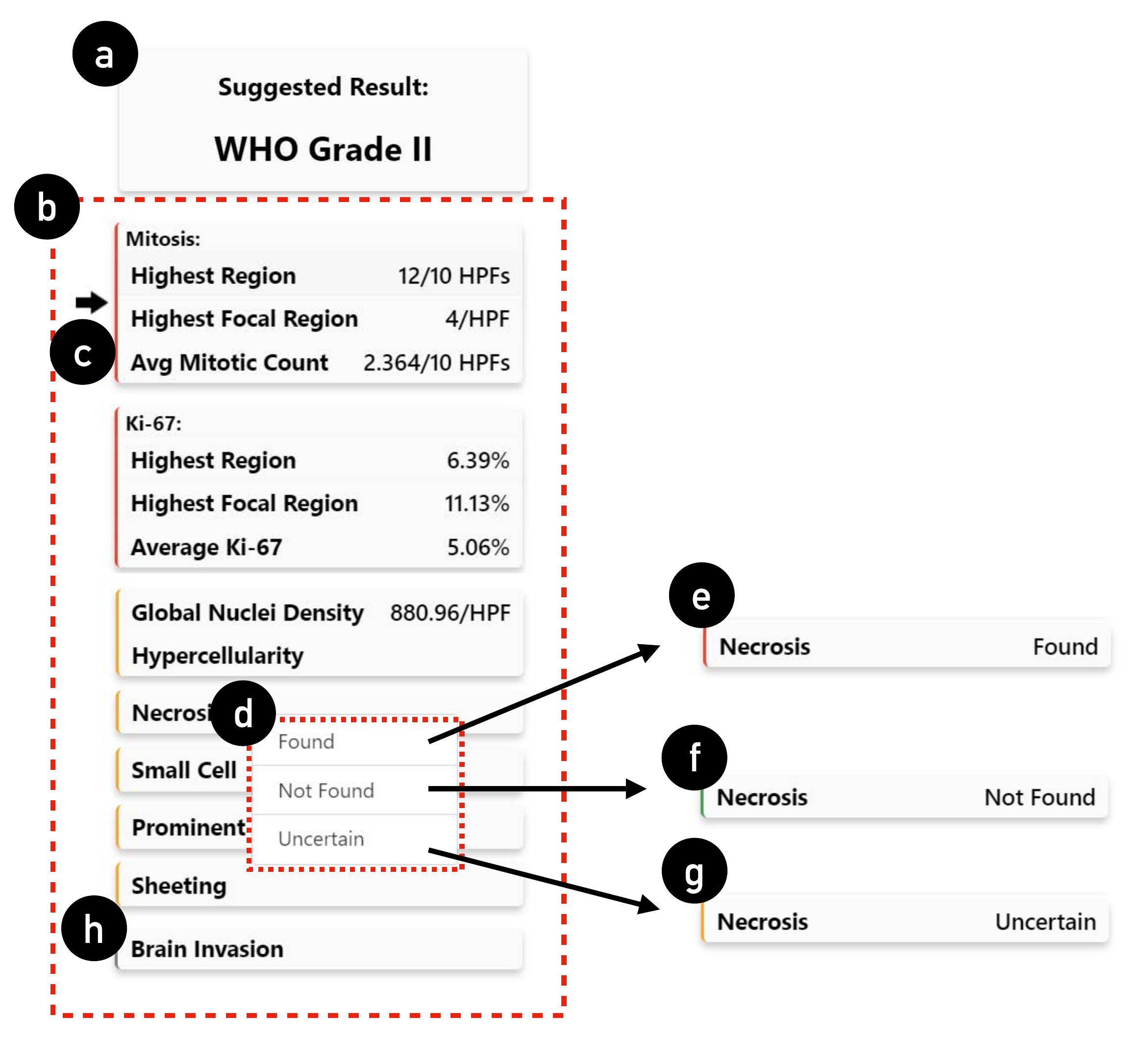}
    \caption{Joint-analyses of multiple criteria in \xp's design: (a) the overall suggested grading; (b) a structured overview of each WHO criterion with  (c) an arrow highlighting the main contributing criterion to the suggested grading; (d) users can override criteria by right-clicking on each item and change the result to `found', `not found' or `uncertain'; \xp~provides color bars to indicate the status of each criterion: (e) red indicates a confirmed abnormal criterion (or \textit{presence}), (f) green indicates a confirmed normal criterion (or \textit{absence}), (g) orange indicates the criterion is unconfirmed/confirmed uncertain, and (h) gray indicates the criterion is not applicable in this case.}
    \label{fig:high_level}
\end{figure}

\subsection{Joint-Analyses of Multiple Criteria}
Based on the formative study, we found that pathologists rely on the WHO meningioma grading guideline for meningioma grading \cite{louis20072007} involving multiple criteria.  Thus \xp's design follows the WHO guideline and employs AI to compute eight critical criteria for meningioma grading\footnote{... which includes the mitotic count, Ki-67 proliferation index, hypercellularity, necrosis, small cell, prominent nucleoli, sheeting, and brain invasion. Note that this work does not consider using AI to identify the subtypes (\eg clear cell, frank anaplasia) because we believe they are relatively easier to be discovered and judged by pathologists.}.
Details on the AI implementation are described in Section \ref{AI_backend}. These criteria can be split into two categories: quantitative and qualitative. For the quantitative criteria (\ie mitotic count, Ki-67 proliferation index), we show their predicted \textit{quantitative values} directly. 
For the other criteria dealing with the \textit{presence} or \textit{absence} of a specific pathological pattern, \xp~ provides recommendations of regions of interest (ROI) hotspots according to the largest aggregations of AIs' probabilities. 

Figure \ref{fig:high_level} demonstrates the interface of multiple criteria, which shows the current suggested grading for the tumor (\ie the suggested `WHO grade 2', Figure \ref{fig:high_level}a) and a structured overview of each criterion (Figure \ref{fig:high_level}b). \xp~ displays an arrow to indicate the main contributing criterion (Figure \ref{fig:high_level}c), the most deterministic AI findings for the suggested diagnosis, according to the meningioma grading guidelines (see Appendix \ref{sec:mg-grade}). For example, in Figure \ref{fig:high_level}, \xp~suggests the ``\textit{mitotic count}'' is the main contributing criterion, because it has detected 12 mitoses in 10 high-power fields (HPFs) (Figure \ref{fig:high_level}c, highest region). Such AI findings directly satisfy descriptions of WHO grade 2 meningiomas, making ``\textit{mitotic count}'' the main contributing criterion. Going beyond the main contributing criterion, all the criteria are linked with the evidence or regions of interest related to the findings. Moreover, AI's recommendation on all the criteria can be overridden by the pathologist (Figure \ref{fig:high_level}d). And \xp~ uses color bars (Figure \ref{fig:high_level}e,f,g,h) to indicate the status.

In summary, the joint-analyses of multiple criteria addresses the challenge of comprehensiveness by providing important information for pathologists according to the medical guideline. \xp~also achieves global explainability by presenting how different AI-computed criteria are combined to arrive at a diagnosis. Such a design can enhance AI's workflow integration because it exposes the pathologist to high-level AI findings when they onboard the case. As such, they can establish an initial understanding and develop hypotheses, which also facilitates them to double-check with their examination later.

\begin{figure}
    \centering
    \includegraphics[width=1.0\linewidth]{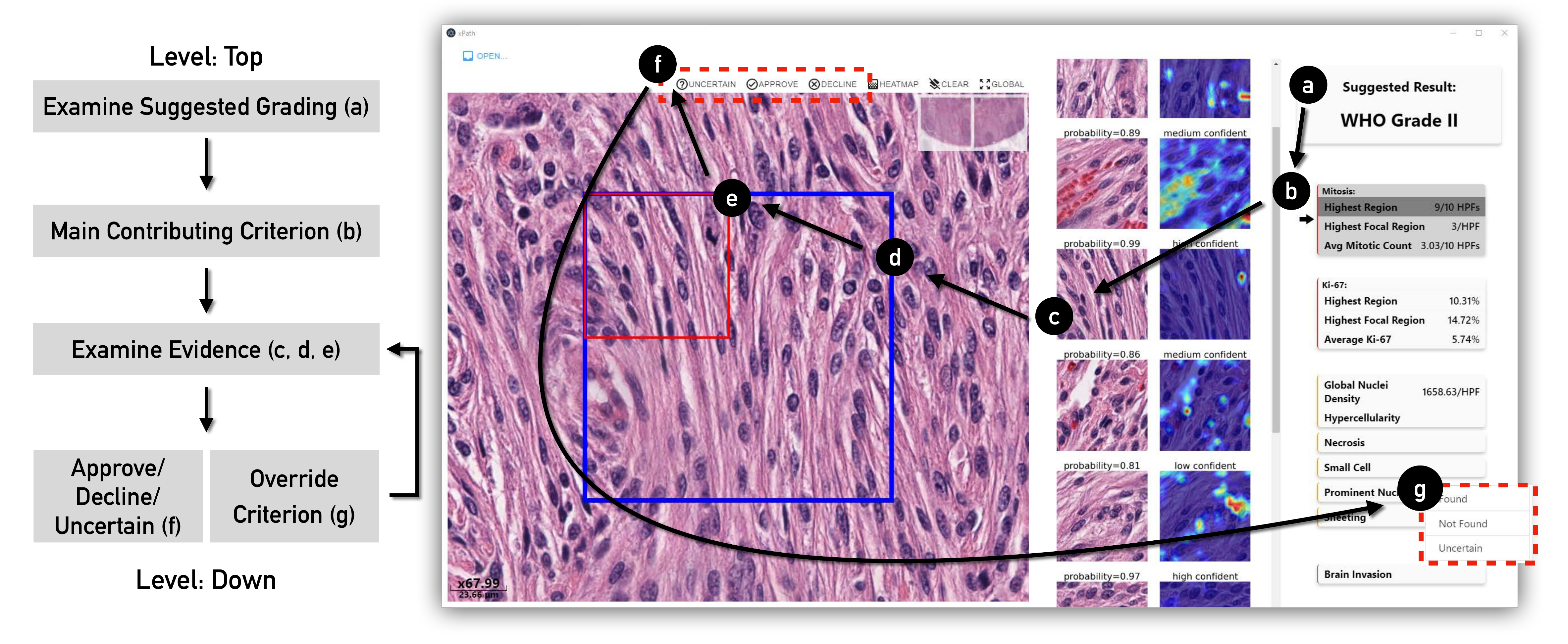}
    \caption{\xp~presents a top-down human-AI collaboration workflow for pathologists to interact with \xp~ (left) and pathologists' corresponding footprints on the \xp's frontend user interface with examining the mitosis criterion as an example (right). A pathologist user starts from (a) the AI-suggested grading result and then examines (b) the main contributing criterion. They can further examine (c) the evidence list, and register back into the original whole slide image in higher magnifications (d,e). Furthermore, users can (f) approve/decline/declare-uncertain on the evidence, or (g) override AI results directly by right-clicking on each criterion. Users might repeat the same workflow (c-g) multiple times to examine other criteria (one criterion for each time). Meanwhile, \xp's suggested grading (a) will be updated as the user justifies AI's findings. The user may continue to interact with \xp~until they have collected sufficient confidence for a diagnosis.}
    \label{fig:work_path}
\end{figure}

\subsection{Explanation by Hierarchically Traceable Evidence for Each Criterion}
\label{evidence_rule}
\begin{figure}
    \centering
    \includegraphics[width=0.4\linewidth]{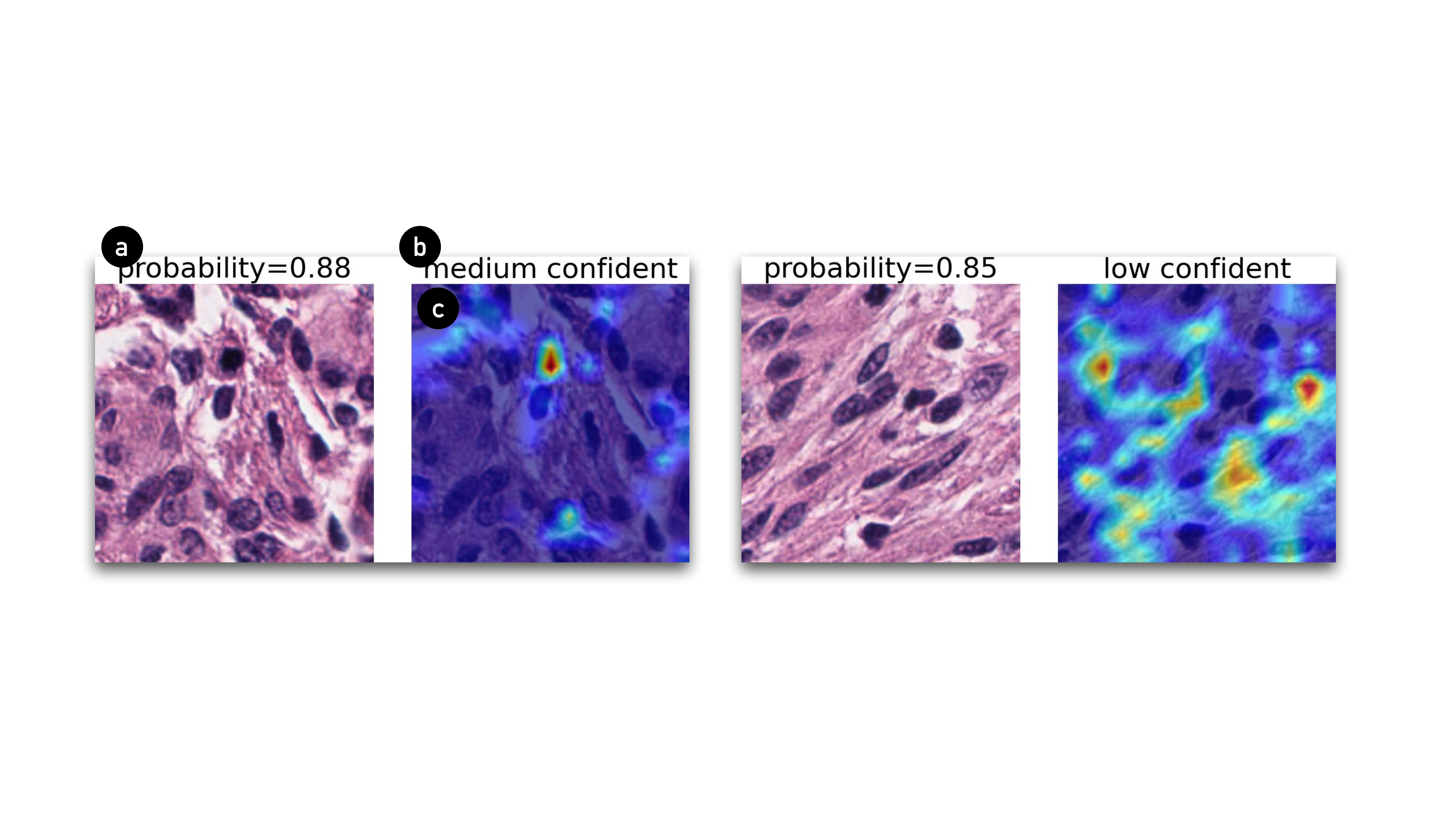}
    \caption{For the mitosis criterion, \xp~demonstrates a series of explanations in each mid-level sample, including the (a) AI's probability, (b) AI's confidence level, which is calculated by the probability thresholds, and (c) a saliency map (calculated by the Grad-CAM++ algorithm \cite{chattopadhay2018grad}) that highlights the spatial support for the mitosis class in the reference image on the left.}
    \label{fig:evidence}
\end{figure}

Another finding from the formative study is that, besides a global explanation of the overall grading, pathologists also would like to see evidence that justifies AI's grading, \eg how AI processes the image of a local patch (for local explainability). Hence, we designed \xp~ to provide such explanations by hierarchically traceable evidence: \xp~enables pathologist users to examine and justify the evidence with a top-down human-AI collaboration workflow. Specifically, at the \textbf{top level}, pathologists can first see the suggested diagnosis recommended by \xp~(Figure \ref{fig:work_path}a). Then, they can continue to dive down and examine a list of AI-computed criteria (Figure \ref{fig:work_path}b). Each criterion can be boiled down to a list of \textbf{mid-level} samples (Figure \ref{fig:work_path}c).  For the most important criterion --- mitosis, \xp~demonstrates a series of explanations in each sample, including AI's output probability (Figure \ref{fig:evidence}a), AI's confidence level (Figure \ref{fig:evidence}b), and a saliency map (Figure \ref{fig:evidence}c) that highlights the spatial support for the mitosis class in the reference image\footnote{Please refer to the supplementary material for the implementation of calculating the confidence level and the saliency map.}, allowing pathologists to check AI's validity on each sample quickly.
Further, at the \textbf{low-level}, \xp~ supports registering each sample into the whole slide image (WSI) to enable pathologists to examine with higher magnification and search nearby for more contextual information (Figure \ref{fig:work_path}d,e). 

With the provided \textbf{mid-} and \textbf{low-level} information, a pathologist can approve/decline/declare-uncertain a sample for a criterion with one click (Figure \ref{fig:work_path}f), or directly override AI's results on each criterion (Figure \ref{fig:work_path}g). Correspondingly, the overall suggested grading (Figure \ref{fig:work_path}a) is updated dynamically upon the user's input. Such a diagnosis-contesting workflow allows pathologists to challenge AI's suggested diagnosis by seeing AI's reasoning line and evidence, which increases the ``contestability'' as described in previous HCI research in healthcare \cite{hirsch2017design}.

Such a workflow mimics a scenario that we found in the formative study: pathologists might assign low-level tasks (\eg marking ROIs, finding specific criteria) to trainees in practice. They can continue to perform a differential diagnosis (\ie building hypotheses and ruling out less-probable cases with findings) based on trainees' reports. By replacing trainees with AI, we emulated the relationship between the pathologists and trainees, thus making AI integral to pathologists' current practices.

\begin{figure}
    \centering
    \includegraphics[width=1.0\linewidth]{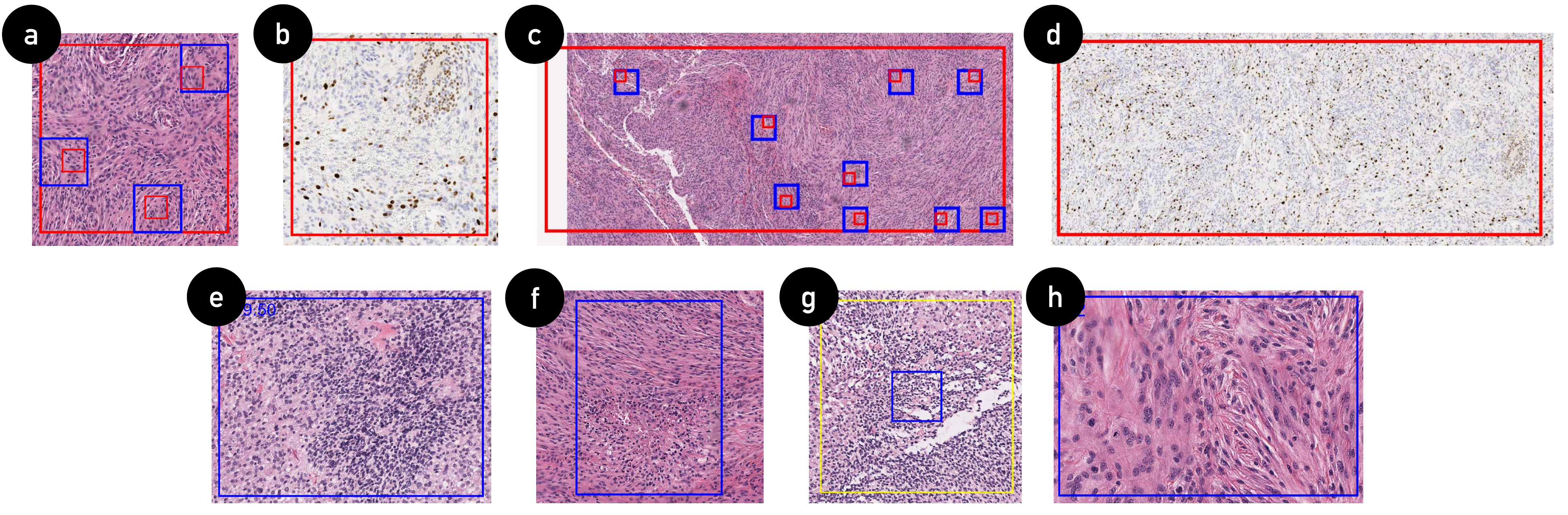}
    \caption{Selected pieces of sampled evidence:
(a) a highest focal region sampling result of mitotic count on H\&E slide (red box, 1HPF), the small blue frames indicate the rough positions of detected mitoses, and the smaller red boxes in the blue frames mark the positions of mitoses (that are shown on the evidence list) found by \xp's AI;
(b) a highest focal region sampling result on the Ki-67 IHC slide (red box, 1HPF); 
(c) a highest region sampling result of mitotic count on H\&E slide (red box, 10HPFs) with mitoses reported by \xp's AI (the blue frames and smaller red boxes);
(d) a highest region sampling result on the Ki-67 IHC slide (red box, 10HPFs);
(e) a hypercellularity ROI sample (blue box);
(f) a necrosis ROI sample (blue box);
(g) a small cell ROI sample (the inner blue box, the outer yellow box marks the dimension of 1HPF);
(h) a prominent nucleoli ROI sample (blue box).}
    \label{fig:hpf_roi_sample}
\end{figure}

Figure \ref{fig:hpf_roi_sample} demonstrates typical examples of evidence provided by \xp. Particularly, for the mitosis-related criteria (\ie mitotic count from H\&E WSI and Ki-67 proliferation index from Ki-67 IHC WSI), which are commonly used for meningioma grading, we introduce two `shortcuts' for pathologists to look into AI's results:

\begin{itemize}
    \item \textbf{Highest Region Sampling}. 
    One WHO criterion is the mitotic count in 10 consecutive high-power fields (HPFs). Our formative study found that the inter-observer consistency of ``10 consecutive HPFs'' is low due to the difference in the ROI sampling rules adopted by pathologists.  To address this problem, \xp~provides the highest region sampling tool. The highest region is defined as a $2\times 5$ HPF area with the highest number of mitotic counts (Figure \ref{fig:hpf_roi_sample}c) or the highest Ki-67 proliferation index (Figure \ref{fig:hpf_roi_sample}d). 
    This tool speeds up a pathologist's work by helping them locate 10 consecutive HPFs as required by the WHO guidelines.
\end{itemize}

\begin{itemize}
    \item \textbf{Highest Focal Region Sampling}. 
    From our formative study, pathologists mentioned that high-grade meningiomas share a common feature of increased mitotic activities in a localized area. Hence, \xp~ provides the highest focal sampling tool to help pathologists better localize highly concentrated mitosis/Ki-67 proliferation index areas. In \xp, the highest focal region is calculated as the one HPF with the highest number of mitotic counts (Figure \ref{fig:hpf_roi_sample}a) or the highest Ki-67 proliferation index (Figure \ref{fig:hpf_roi_sample}b). Using this tool, pathologists can locate foci of highly-mitotic areas that the highest region sampling might miss.
\end{itemize}

Pathologists can go beyond the sampled areas and navigate the high-heat areas using heatmaps generated for the whole slide (please see the supplementary material for details). For example, the mitosis heatmap registers all AI-detected positive mitotic cells as a mitotic density atlas, where high-heat areas indicate a high density of mitotic cells. As such, the heatmap would serve as a `screening tool' to help pathologists filter out unrelated areas and rapidly narrow down to the ROIs that are scattered in an entire WSI. \xp~ provides such `screening tools' for all criteria. 

After pathologists have finished examining one criterion, they can proceed to justify the rest of the criteria with the same top-down workflow (one iteration for each criterion). During such an iterative process, \xp~will update AI's findings on an individual criterion and, if necessary, the overall suggested grading as well. Finally, pathologists can make a diagnosis once they have collected sufficient confidence for the grading diagnosis.

In summary, in contrast to prior work that enables pathologists to define their own criteria for finding similar examples \cite{cai2019human}, \xp~aims at making examinations based on an existing criterion traceable and transparent with evidence, which allows pathologists to see and understand why AI derives such findings. Furthermore, pathologists can challenge (or ``contest'' \cite{hirsch2017design}) these AI findings with a top-down workflow to refine the suggested grading diagnosis. Such collaboration between pathologists and AI is similar to that with pathology trainees, where pathologists can perform a differential diagnosis based on trainees' findings.

%% file: 06_implementation.tex
\section{Implementation of xPath's AI Backend}
\label{AI_backend}

\xp~implements an AI-aided pathology image processing backend to compute the eight pathological criteria of the mitotic count, Ki-67 proliferation index, hypercellularity, necrosis, small cell, prominent nucleoli, sheeting, and brain invasion. 
In this section, we briefly describe datasets, the AI processing pipeline, and AI training details.
Finally, we report the performance of each of the AI models from a technical evaluation.

\begin{figure}
    \centering
    \includegraphics[width=1\linewidth]{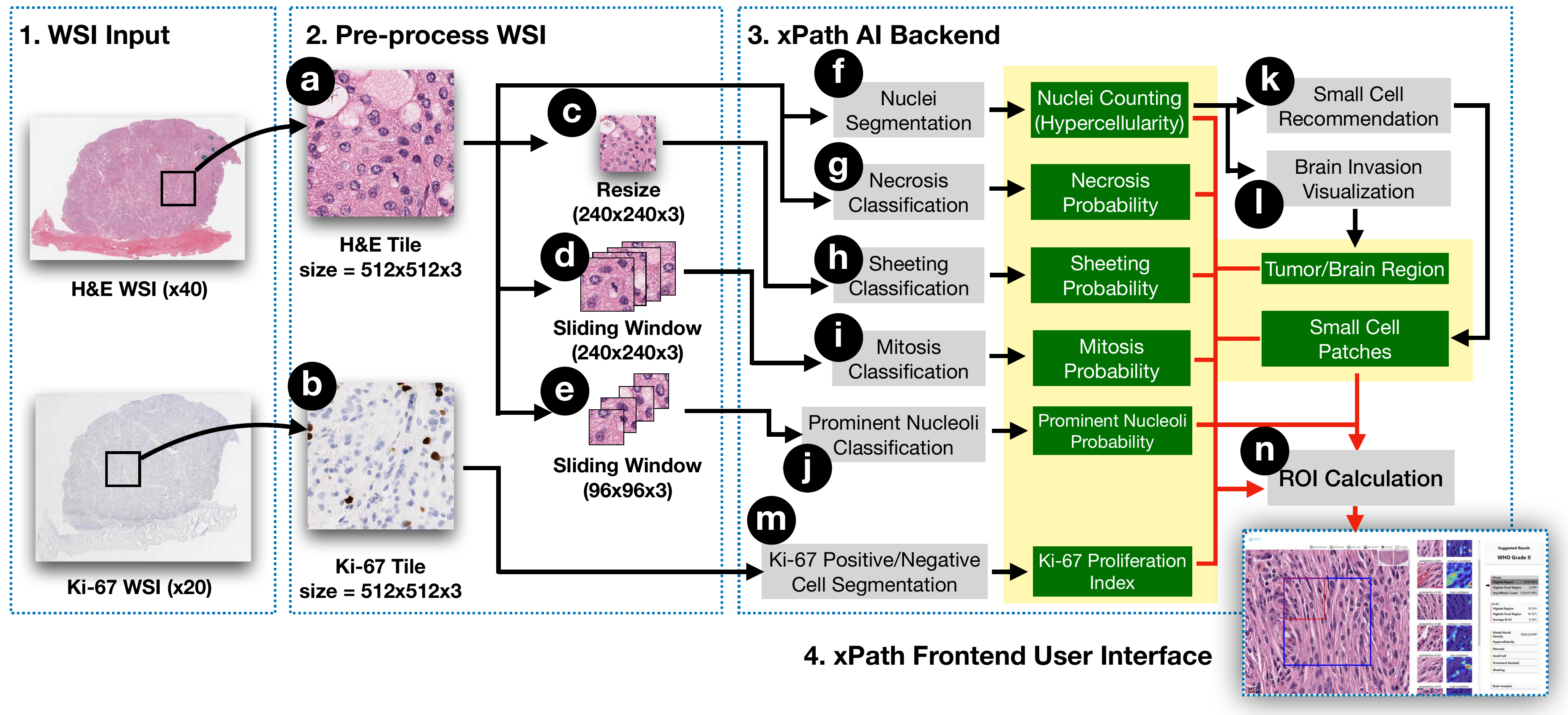}
    \caption{Data processing pipeline of \xp: \one \xp~takes H\&E and Ki-67 whole slide images (WSIs) as input. \two For each WSI, \xp~uses a sliding window method to acquire (a) H\&E and (b) Ki-67 tiles; Furthermore, each H\&E tile is processed with (c) resizing, (d) sliding window ($240\times240\times3$), and (c) another sliding window ($96\times96\times3$) to fit the inputs of the down-stream AI models. \three \xp's AI backend takes over the pre-processed tiles and employs multiple AI models to detect WHO meningioma grading criteria from each tile. Given an H\&E tile, \xp~uses (f) a nuclei segmentation model to count the number of nuclei (for hypercellularity judgment), (g) a necrosis classification model to calculate necrosis probability, and (h) a sheeting classification model to calculate sheeting probability. \xp~further utilizes the nuclei counting results for (k) small cell recommendation, and (l) brain invasion visualization. For a $240\times 240 \times 3$ tile,  \xp~uses (i) a mitosis classification model to obtain the mitosis probability. For a $96\times 96 \times 3$ tile,  \xp~uses (j) a prominent nucleoli classification model to predict prominent nuclei probability. For each Ki-67 tile, \xp~(m) detects positive and negative nucleus to calculate the Ki-67 scores; \four \xp~further (n) calculates ROIs based on all AI-computed results (marked in the green boxes), and shows them as evidence on the frontend user interface for pathologist users to justify.}
    \label{fig:backend_diagram}
\end{figure}

\subsection{Processing WSIs with AI}
\xp~aims to screen the entire whole slide image (WSI) using AI and then determine suggested grades based on the AI findings. To achieve this, \xp~ includes six AI models and two rules, one for each criterion, to general initial AI results.
For each WSI, we first used a sliding window technique to cut it into smaller tiles. For each tile, we further employed a series of AI models to calculate six criteria (\ie nuclei count (Figure \ref{fig:backend_diagram}f), necrosis probability (Figure \ref{fig:backend_diagram}g), sheeting probability (Figure \ref{fig:backend_diagram}h), mitosis (Figure \ref{fig:backend_diagram}i), prominent nucleoli (Figure \ref{fig:backend_diagram}j), and Ki-67 proliferation index (Figure \ref{fig:backend_diagram}m)). Based on the AI-computed nuclei count, we further used two rules to support the reporting of the small cell and the brain invasion patterns. \xp~can recommend small cell tiles based on the nuclei count of each tile (Figure \ref{fig:backend_diagram}k). Furthermore, the brain invasion was visualized by classifying the brain \textit{vs.} tumor regions according to the nuclei count (Figure \ref{fig:backend_diagram}l). This is because meningioma tumor areas usually have a high nuclei density, while normal brain tissues are not. After the AI models had processed each tile,  \xp~calculated the ROIs using a set of rules. Please refer to the supplementary material for more detailed descriptions of \xp's AI implementation and the ROI generation process.

\subsection{Dataset and Model Training}

Since there were no pre-trained models nor public meningioma datasets for the pathology patterns of mitosis, necrosis, prominent nucleoli, and sheeting, we built an in-house dataset consisting of 30 WSIs (WSI total size = $\sim$ 54.9 GB) from a local medical center to train AI models to classify these four patterns. The WSIs were scanned by an Aperio CS2 scanner in x400 magnification (pixel size=0.25$\mu$m). The ground truth labels were collected in two ways: \one for the mitosis, the pathologist labeled with an online labeling system; \two for other criteria, the pathologist marked ROIs using the Imagescope software. We then cropped the labeled ROIs with a random-crop technique, and the tiles in different sets were generated from a different group of ROIs. In sum, the final dataset has a size of $\sim$ 16.1 GB. It consists of four training and testing sets, covering the four pathology patterns (as shown in Table \ref{tab_dataset}).

To train the models, for each criterion, we further randomly selected a subset of the training set to be the validation set. Specific thresholds were decided by the maximum F1 scores achieved by each model in the validation set. Please find the supplementary material for more specific training details.

\begin{table*}
    \footnotesize
\centering
\begin{tabular}{c | c c c}
\hline
Dataset & \begin{tabular}[x]{@{}c@{}}Dimension\\ (in pixels) \end{tabular} & \begin{tabular}[x]{@{}c@{}}\# of Samples\\ (Training) \end{tabular} &  \begin{tabular}[x]{@{}c@{}}\# of Samples\\ (Testing) \end{tabular}\\
\hline \hline
Mitosis & $256\times256\times3$ & 33,562 (1,925 positive, 31,637 negative) & 8,223 (336 positive, 7,887 negative)\\
\hline
Necrosis & $512\times512\times3$ & \begin{tabular}[x]{@{}c@{}}4,383 (from 190 regions) \\ (651 positive, 3,732 negative) \end{tabular} & \begin{tabular}[x]{@{}c@{}} 3,587 (from 162 regions) \\(770 positive, 2,817 negative) \end{tabular}\\
\hline
Prominent Nucleoli & $96\times96\times3$ & 15,042 (2,447 positive, 12,595 negative) & 3,753 (609 positive, 3,144 negative)\\
\hline
Sheeting & $240\times240\times3$ & \begin{tabular}[x]{@{}c@{}}3,660 (from 55 regions) \\ (1605 positive, 2055 negative) \end{tabular} & \begin{tabular}[x]{@{}c@{}}2,340 (from 45 regions)\\ (1,185 positive, 1,155 negative) \end{tabular}\\
\hline
\end{tabular}
\caption{The description of the dataset for each task. The dimensions of input tiles (in pixels), the size of training/testing sets, and the distribution of positive/negative tiles are provided. }
\label{tab_dataset}
\end{table*}

\subsection{Technical Evaluation}
\label{sec:tech_eval}
\begin{figure}
    \centering
    \includegraphics[width=1.0\linewidth]{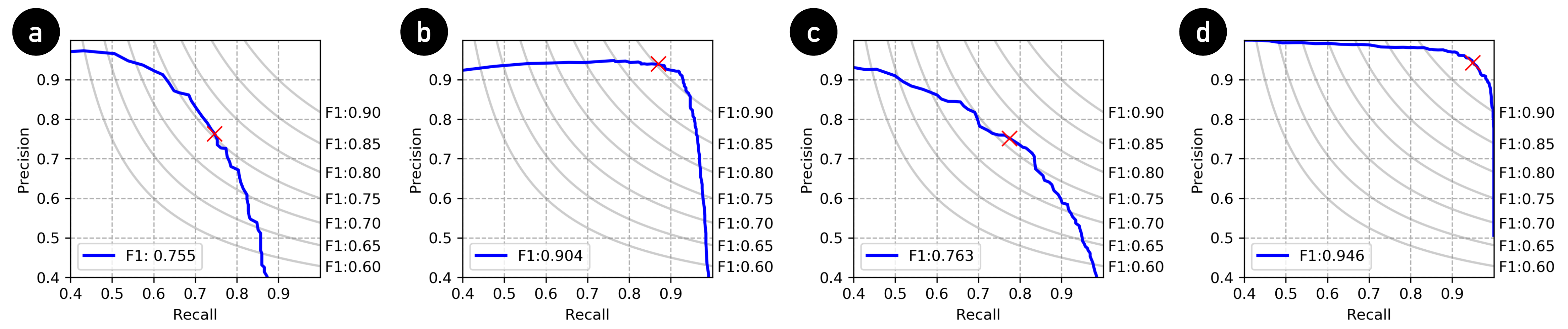}
    \caption{Classification performance for (a) mitosis, (b) necrosis, (c) prominent nucleoli, (d) sheeting. The solid blue lines in each sub-figure illustrate the Precision-Recall curves of each model. The red crosses indicate the performance achieved by the models using the thresholds that maximized the F1 scores on the validation sets. The gray lines in each figure are the height lines of the F1 scores. The F1 score of each height line is shown on the right axis.}
    \label{fig:performance}
\end{figure}

We report the performance of AI models on testing sets. Specifically, we test the supervised models for recognizing mitosis, necrosis, prominent nucleoli, and sheeting criteria, and report the Precision-Recall curve, as shown in Figure \ref{fig:performance}. In summary, \xp~ achieved F1 scores of 0.755, 0.904, 0.763, and 0.946 in identifying the pathological patterns of mitosis, necrosis, prominent nucleolus, and sheeting. The scores indicate the effectiveness of our models. Moreover, for the tasks of cell-counting in hypercellularity and Ki-67 proliferation index criteria, we test their performance with 150 randomly-selected $512\times 512\times 3$ tiles each and report the average error rate. The results show that the average error rate of nuclei counting (hypercellularity) and Ki-67 proliferation index is 12.08\% and 29.36\%, respectively.

Due to a lack of data at present, for brain invasion and small cell patterns, rather than drawing a definitive conclusion, \xp~ uses a rule-based, unsupervised approach to recommend areas for pathologists to examine. We planned to validate the performance on these two criteria later in the work sessions with pathologists; however, it was hard for the participants to differentiate the small cell formation {\it vs.} inflammation areas without proper IHC tests. As such, \xp's AI performance in detecting small cell patterns was not validated. For the brain invasion, most pathologists felt it was faster to examine it manually and did not rely on AI's recommendations.

%% file: 07_evaluation.tex
\section{Work Sessions with Pathologists}

The technical evaluation reported in the previous session validated the effectiveness of \xp's AI backend in the in-house dataset. However, it remains unanswered whether \xp~is beneficial to pathologist users in practice. Notably, many previous cases showed how easily AI models could break, although they showed high accuracy in training/test data \cite{strickland2019ibm, kandula2019reappraising}. To address these concerns, we conducted work sessions with 12 medical professionals in pathology across three medical centers and studied their behavior of grading meningiomas using a traditional interface --- an open-source whole slide image viewer called ASAP\footnote{\url{https://computationalpathologygroup.github.io/ASAP/}. This tool was selected because it is open-source and has gained popularity in the digital pathology research domain \cite{litjens20181399}.} and \xp. In this study, we referred to the traditional interface as system 1 and \xp~as system 2 to avoid biasing of participants. The main research questions are:

\textit{RQ1: Can \xp~ enable pathologists to achieve accurate diagnoses?}

One reason for utilizing AI in \xp~is because it can highlight ROIs of multiple pathological patterns, freeing pathologists from examining the entire slide. However, it is still yet unclear whether introducing AI will have a positive or a negative effect on pathologists' diagnoses: On one hand, multiple previous works show that the introduction of human-AI collaboration improves pathologists' performance \cite{wang2016deep, bulten2021artificial}; On the other hand, due to the existing limitations in AI models' accuracy, users face the risk to generate wrong diagnoses if they over-rely on the non-perfect AI \cite{bansal2019beyond, buccinca2021trust}. Since there is no solid conclusion on this, we hypothesize that ---

\begin{itemize}
\item \textbf{[H1] Pathologists' grading decisions with \xp~will be as accurate as those with manual examinations.}
\end{itemize}

\textit{RQ2: Do pathologists work more efficiently with \xp?}

Another reason for using AI in \xp~is that it can improve the pathologists' throughput by alleviating their workload. However, it remains unanswered how AI will assist pathologists in \xp, given that previous work shows less-carefully-designed AI might incur extra burdens \cite{gu2020lessons}. As such, it is also necessary to find out whether pathologists can work efficiently with \xp's AI. We hypothesize that ---

\begin{itemize}
    \item \textbf{[H2a] Pathologists will spend less time examining meningioma cases using \xp.}
    \item \textbf{[H2b] Pathologists will perceive less effort using \xp.}
\end{itemize}

\textit{RQ3: Overall, does \xp~ add value to pathologists' existing workflow?}

Going beyond the influence brought by AI, we introduce two design ingredients for pathology AI systems --- joint-analyses of multiple criteria \textit{and} explanation by hierarchically traceable evidence in \xp. We also concluded three system requirements, \ie comprehensiveness, explainability, and integrability for \xp. In this study, we investigate whether such designs will add value to pathologists' existing workflow. Specifically, we hypothesize that:

\begin{itemize}
    \item \textbf{[H3a] \xp~will improve comprehensiveness with the joint-analyses of multiple criteria.}
    \item \textbf{[H3b] \xp~will improve explainability with explanation by hierarchically traceable evidence.}
    \item \textbf{[H3c] \xp~will improve integrability with the top-down human-AI collaboration workflow.}
\end{itemize}

\subsection{Participants}
We recruited 12 medical professionals in pathology across three medical centers in the United States through word-of-mouth and by sending flyers to the mailing lists. All participants were required to complete at least one year of post-graduate pathology residency training ($\geq$ PGY-2). Our participants' experience ranged from two to ten years ($\mu$=4.38, $\sigma$=2.16), including two attendings (A), two fellows (F), seven senior residents (SR, $\geq$ PGY-3), and one junior resident (JR, PGY-2). The demographic information of the participants is shown in Table \ref{tab_user_work_session}. All participants had received training for examining meningiomas before the work sessions. And all participants had experience in seeing digital pathology slides prior to the study. They primarily used the Imagescope (a commercial software that provides image viewing functions similar to the ASAP) to see whole slide images (WSIs). The primary purpose of using the digital system was to train or review remote cases.

\begin{table}
  \footnotesize
\centering
\begin{tabular}{c | c c c | c c c c c c c}
\hline
 ID & Occupation & \makecell{Years of \\ Experience}  & \makecell{ Frequency of \\Seeing WSIs} & ME194 & ME195 & ME196 & ME197 & ME198 & ME199\\
\hline \hline
 P1 & PGY-3 & 3 & Weekly & & ASAP & xPath & & & \\
 P2 & PGY-4 & 4 & Monthly & ASAP & & xPath & xPath & xPath & xPath\\
 P3 & Fellow & 4 & In Six Months & & & xPath & & ASAP \\
 P4 & Fellow & 5 & Weekly & & xPath & ASAP & & xPath & \\
 P5 & PGY-4 & 4 & Weekly & xPath & xPath & & & ASAP &\\
 P6 & PGY-3 & 3 & Monthly & &  & xPath & ASAP & &\\ 
 P7 & Attending & 7 & Weekly & &  &  & xPath & ASAP & xPath \\ 
 P8 & PGY-4 & 3.5 & Weekly & & xPath & & & & ASAP \\ 
 P9 & PGY-2 & 2 & Bi-weekly & ASAP & & & xPath & & \\ 
 P10 & PGY-3 & 3 & Weekly & xPath & & & ASAP & xPath & \\ 
 P11 & PGY-4 & 4 & Monthly & & ASAP & xPath & & &  \\
 P12 & Attending & 10 & Weekly & & xPath & xPath & & ASAP & \\
\hline
\end{tabular}
\caption{Demographic information \& arrangements of the participants in the work sessions. `ME195' -- `ME199' are the case IDs. During the study, participants used `ASAP' (system 1) and `xPath' (system 2) to examine the cases. Note that FP12 had also participated in the formative study (referred to as FP3 in Table \ref{tab_user}.)}
\label{tab_user_work_session}
\end{table}

\subsection{Test Data}
\label{subsec:test-data}
\begin{figure}
    \centering
    \includegraphics[width=1.0\linewidth]{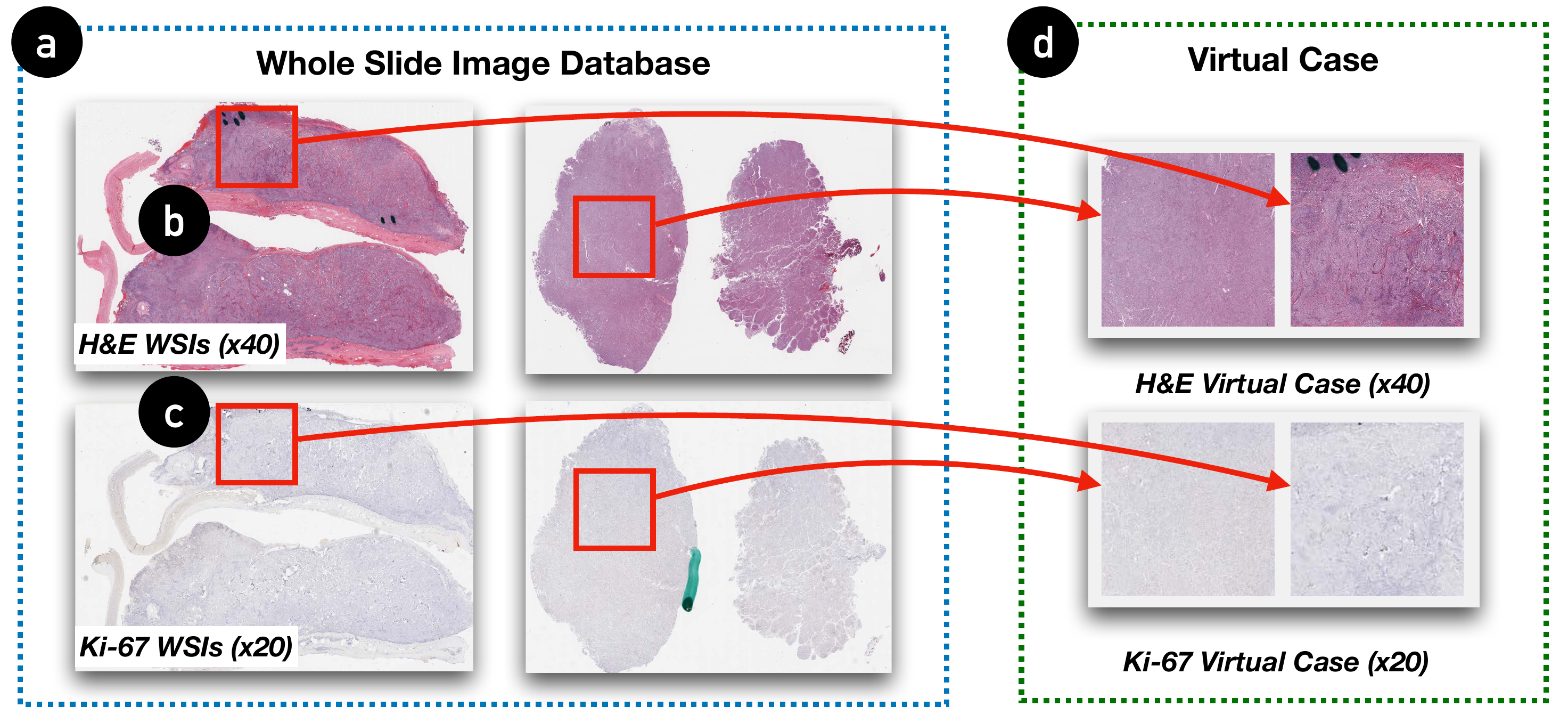}
    \caption{We used the `virtual cookie cut' technique to generate the tests cases. Specifically, we first collected (a) pairs of H\&E (in x400) and Ki-67 (in x200) WSIs. Then, we generated `virtual cuts' by (b) selecting 30,000$\times$30,000-pixel regions in H\&E WSIs, and (c) 15,000$\times$15,000-pixel regions from the same position as their H\&E counterparts. (d) Each virtual case consists of one mandatory H\&E slide with two nodes and one optional Ki-67 slide with two corresponding ones. }
    \label{fig:cookie-cut}
\end{figure}

We asked our pathologist collaborators in a local medical center to select 18 meningioma slides and scan them to WSIs\footnote{\dots which include eleven H\&E WSIs (scanned in x400), and seven Ki-67 WSIs (scanned in x200).} with an Aperio CS2 scanner to generate the test cases (IRB\#20-000431). In normal conditions, each patient's case consisted of more than 10 WSIs, and an averaged-experienced resident pathologist typically needs to spend about one hour to finish examining an averaged-difficult case (\ie criteria found in the case do not lie on the grading borderlines). As such, we generated nine `virtual patient cases' with the `virtual cookie cut' technique (see Figure \ref{fig:cookie-cut}) to fit the task of grading meningiomas in hour-long working sessions.

Each virtual patient consisted of a mandatory H\&E slide (in x400), and an optional Ki-67 slide (in x200). Each H\&E slide had two nodes (each has a size of 30,000$\times$30,000 pixels), while each Ki-67 slide had two corresponding Ki-67 nodes (each has a size of 15,000$\times$15,000 pixels) that were extracted from the same position as their H\&E counterparts, if available. The contours of nodes were removed as a ``wash-out'' measure because some participants had seen the slides before the study. All nodes were selected by an expert pathologist and included deterministic regions of interest (\ie  crucial areas that include necessary information) for the diagnosis. Therefore, although participants were seeing virtual patients in the study, they still had to use the full system to diagnose because pathological criteria in the test data were not eliminated. In total, nine virtual cases have nine H\&E slides and six Ki-67 slides. 

The ground truth diagnoses was provided by an experienced pathologist, including two WHO grade 1, five WHO grade 2, and two WHO grade 3. We selected three from the grade 2 cases for the tutorial purpose, leaving the test set with two cases for each grade.

\subsection{Task \& Procedure}
All sessions were conducted online because of the COVID-19 pandemic. We first introduced the project's mission and provided a detailed walkthrough of the traditional interface and \xp~ with three pairs of H\&E and Ki-67 slides as an example. Participants used Microsoft Remote Desktop to interact with both systems that ran on a remote server. Next, we ran a testing session for the participants to grade one virtual case with the traditional interface, and one-four others using \xp~with the time cost logged. The variation in the cases was caused by the between-subject difference in the time consumption of using \xp. And such a difference was caused by two factors: \one participants' learning abilities --- some learned faster to use \xp~than others; \two participants' abilities in examining the evidence. The order was counterbalanced across participants. 

For each case, the time was counted from when participants first clicked the WSI case until they reached the grading diagnosis. After participants finished each case, we asked them to report their grading diagnosis as well as their findings through a questionnaire adapted from the College of American Pathologists (CAP) cancer protocol template\footnote{\url{https://documents.cap.org/protocols/cp-cns-18protocol-4000.pdf}}. In this session, we did not compare \xp~ with traditional optical microscopes because of the difficulty of instrumentation and observation given the remote situation.
After participants had examined all the cases, we conducted a semi-structured interview to elicit their responses to \xp's perceived effort and added value. The average duration of each work session was $\sim$70 minutes. Although conducted online, we set up the testing environment as close to pathologists' everyday clinical workflow: \one we used H\&E and Ki-67 data based on real patients (as described in Section \ref{subsec:test-data}); \two we used real working systems of ASAP and \xp; \three we asked our participants to diagnose following the same examination protocol as they had done in practice.

\subsection{Measurements}
In this study, we collected participants' grading decisions from the CAP questionnaire and analyzed the time log. We also asked them to fill in a post-study questionnaire (see Table \ref{tab:study_quant}) with seven-point Likert questions following \cite{cai2019human, jordan1996usability, hart1988development}. We tested our hypotheses via the following measurements:

For \textbf{H1}, we compared the diagnoses reported by participants and the ground-truth diagnoses. We measured the accuracy of both systems by calculating the error rates.

For \textbf{H2a}, we calculated the average time participants spent on each case using \xp~and the traditional interface. For \textbf{H2b}, we asked them to give both systems ratings of the effort needed for grading (Table \ref{tab:study_quant}, W1), and the effectiveness of the system in reducing the workload (Table \ref{tab:study_quant}, W2) in the post-study questionnaire.

{\bf H3a-c} was evaluated by the post-study questionnaire. For \textbf{H3a}, we asked participants to rate the comprehensiveness of \xp~and the traditional interface (Table  \ref{tab:study_quant}, C1).  For \textbf{H3b}, we asked them to rate the explainability of \xp~only since the traditional interface did not provide AI detections (Table \ref{tab:study_quant}, E1). For \textbf{H3c}, we asked participants to rate the integrability of both systems (Table \ref{tab:study_quant}, I1). Because ``comprehensiveness'', ``explainability'' and ``integrability'' are non-trivial terms, we included the following clarifications for the three terms in the questionnaire: 

\begin{itemize}
    \item \textbf{``Comprehensiveness''}: ``\textit{whether the system can provide detections for (1) multiple criteria for diagnosis and (2) entire slide, instead of a local area;}''
    \item \textbf{``Explainability''}: ``\textit{(1) how results from multiple criteria are combined to yield a grading; (2) what evidence leads to the value of each criterion; (3) why AI thinks a piece of evidence is positive / negative;}''
    \item \textbf{``Integrability''}: ``\textit{whether the system is integrable to your workflow of examining meningiomas.}''
\end{itemize}

Apart from the hypotheses, we also asked the participants to rate the helpfulness of each component in \xp~(``\textit{Rate the helpfulness of each component.}'' --- 1=lowest and 7=highest). Next, we investigated whether the participants trusted \xp~by asking them the following two questions: \one \textit{How capable is the system at helping grade meningiomas?} (Table \ref{tab:study_quant}, T1), \two \textit{How confident do you feel about the accuracy of your diagnoses using the system?} (Table \ref{tab:study_quant}, T2). Last but not least, to evaluate participants' attitudes towards \xp's workflow integration, we asked whether the participants would like to use both systems in the future (Table \ref{tab:study_quant}, F1), and also let the participants rate the overall preference of system 1 \textit{vs.} system 2 (Table \ref{tab:study_quant}, F2).

%% file: 08_finding.tex
\section{Results \& Findings}

In this section, we first discuss our initial research questions and hypotheses. Then, we summarize the recurring themes that we have found in the working sessions.

\subsection{RQ1: Can \xp~ enable pathologists to achieve accurate diagnoses?}
We summarize the CAP questionnaire responses from our participants and collect 12 grading decisions from the traditional interface and 20 from \xp.  We then follow previous works on digital pathology \cite{tschandl2020human, steiner2020evaluation} and compare the difference between participants' responses and the ground truth diagnoses. In summary, with the traditional interface, participants gave correct grading decisions for 7/12 cases, lower-than-ground-truth gradings for 4/12 cases, and higher-than-ground-truth grading for 1/12 cases. In comparison, using \xp, participants gave 17/20 cases correct gradings and lower-than-ground-truth gradings in 3/20 diagnoses. Upon further analysis, we found that all three errors that participants made with \xp~were caused by their over-reliance on AI. In these cases specifically, participants spent the majority of their effort examining the evidence reported by \xp~and missed the false-negative features that \xp~failed to detect --- \quo{ It's just that I got caught up in looking at the boxes, and I would forget that I should look at the entire case myself.}{4}

In sum, based on the data collected by the study, we report that participants could make more accurate grading decisions with \xp~compared to the traditional interface (\textbf{H1}).

\subsection{RQ2: Do pathologists work more efficiently with \xp?}

Contrary to our hypothesis (\textbf{H2a}), participants spent an average of 7min13s examining each case using \xp, which is 1min17s higher than the traditional interface (ASAP). Our study suggests that participants tended to ($p$=0.050, Wilcoxon rank-sum test, same below) invest more time in \xp~than the traditional interface. We believe this is partly because \xp~brings participants an extra workload to comprehend and justify the AI findings. In the traditional interface, our participants share a similar workflow of examining the WSI --- they first scanned the entire WSI in low magnification, then prioritized studying one criterion (such as the brain invasion or the mitotic count) to ascertain a probable diagnosis as quickly as possible. They also checked Ki-67 slides to support their diagnosis. In this process, they collected evidence that accounts for a higher grade and memorized them in their minds. Once they acquired enough evidence, they would stop and make a grading decision. 
When using \xp, participants did not abandon their standard workflow as in the traditional interface. Rather, on top of their standard workflow, participants would perform the differential diagnosis based on AI's findings --- they clicked through each piece of evidence in \xp, justified it by registering into the WSI, and at times overrode AI by clicking the approve/decline/declare-uncertain buttons. These extra steps of interactions prolong participants' workflow ---

\quo{System 2 (\xp) actually makes it longer because some of the images have sort of competing opinions --- whether this is mitosis or not \dots So I'd better take a closer look at what the machine suggests.}{3}

Regarding the perceived effort (\textbf{H2b}), participants reported significantly less effort (Table \ref{tab:study_quant}, W1, \xp: $\mu$=0.91, ASAP: $\mu$=3.67, $p$=0.002) and a stronger effect on reducing the workload (Table \ref{tab:study_quant}, W2, \xp: $\mu$=5.83, ASAP: $\mu$=2.17, $p$=0.002) while using \xp. Participants mentioned that automating the process of finding small-scaled histopathological features, especially mitosis, would save their time and effort ---

\quo{I spend a lot more time crawling around the slide in the high-power, looking for mitosis (for system 1), which you don't have to do as much in system 2 (\xp).}{8}

\begin{table}
    \scalebox{0.84}{
        \begin{tabular}{l|cc}
        \hline 
            \textbf{Questions} & \textbf{ASAP} & \textbf{\xp}\\
        \hline\hline
        
        C1: Rate the comprehensiveness of the system. & 2.83(1.27) & 5.75(0.75)\\
        E1: Rate the explainability of the system. & N/A & 5.58(0.90)\\
        I1: Rate the integrability of the system. & 4.17(1.70) & 5.91(1.08)\\
        W1: Rate the effort needed to grade meningiomas when using the system. & 3.67(1.37) & 0.91(0.90)\\
        W2: Rate the effect of the system on your workload to reach a diagnosis. & 2.17(1.40)& 5.83(1.03)\\
        T1: How capable is the system at helping grade meningiomas? & N/A & 5.83(0.94) \\
        T2: How confident do you feel about the accuracy of your diagnoses using the system? & N/A & 6.00 (0.95)\\
        F1: If approved by the FDA, I would like to use this system in the future. & 3.75(1.76) & 6.42(0.79)\\
        F2: Overall preference & \multicolumn{2}{c}{6.75(0.45)} \\
        \hline
        \end{tabular}
        }
        
        \caption{Participants' response of average scores (and standard deviation) on the quantitative measurements of a traditional interface (ASAP) and \xp~with seven-point Likert questions. For the rating questions (C1, E1, I1, W1, W2), 1=lowest and 7=highest. For question T1, T2, F1, 1=very strongly disagree, 2=strongly disagree, 3=slightly disagree, 4=neutral, \dots,  and 7=very strongly agree. For question F2, 1=totally prefer system 1 over system 2, 2=much more prefer system 1 over system 2, 3=slightly prefer system 1 over system 2, 4=neutral, \dots, and 7=totally prefer system 2 over system 1. Note that for question W1, a higher score indicates that users perceive more effort while using the system. Question E1, T1, T2 are not applicable to ASAP, since it does not provide AI assistance.
        }
        \label{tab:study_quant}
\end{table}

\subsection{RQ3: Overall, does \xp~ add value to pathologists' existing workflow?}
\begin{figure}
    \centering
    \includegraphics[width=1.0\linewidth]{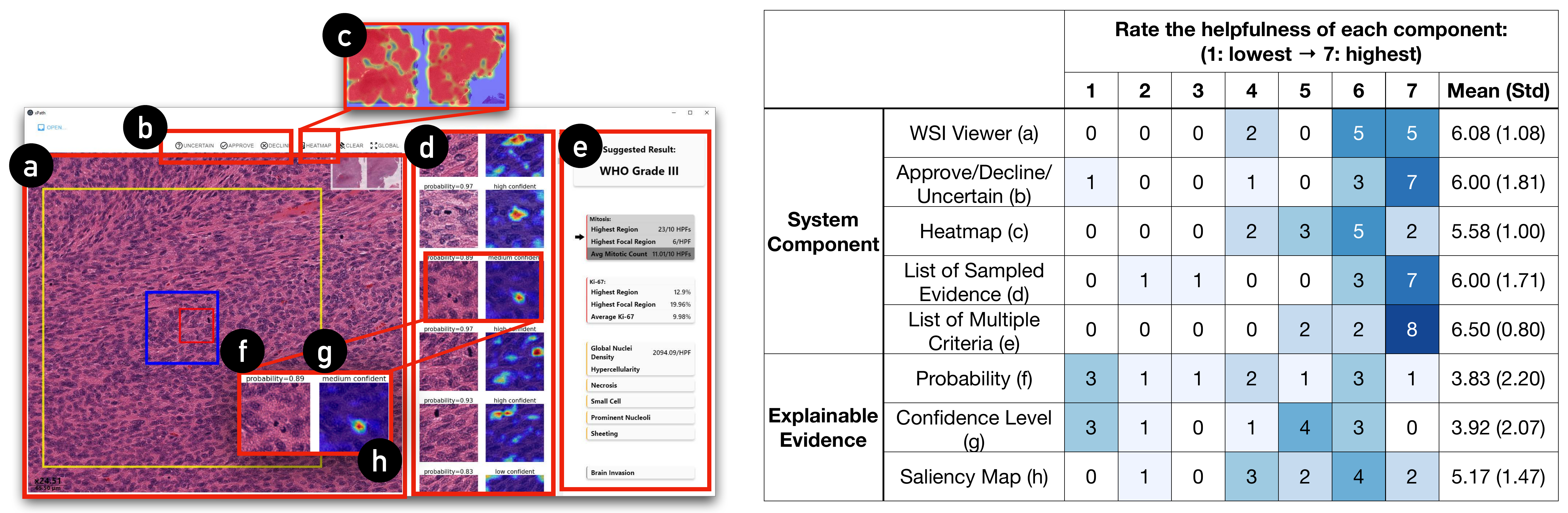}
    \caption{Participants' helpfulness ratings of each component in \xp. Each letter-labeled component in the right table corresponds to the marked part on the left.}
    \label{fig:rating}
\end{figure}

For the comprehensiveness dimension (\textbf{H3a}), \xp~received a significantly higher rating than the traditional interface (Table \ref{tab:study_quant}, C1, \xp: $\mu$=5.75, ASAP: $\mu$=2.83, $p$=0.001). Furthermore, participants gave an average helpfulness score of 6.50/7 for the design of joint-analyses of multiple criteria (see Figure \ref{fig:rating}e). They responded positively that such a design provides sufficient information (\ie criteria and evidence) to assist the diagnosis ---

\quo{\dots it (\xp) kind of gives you a step-wise checklist to make sure that it's the correct diagnosis, and also provides you what is most likely a diagnosis.}{11}

For the explainability dimension (\textbf{H3b}), \xp~obtained an average rating of 5.58/7 (Table \ref{tab:study_quant}, E1). In general, participants could understand the logical relationship between the evidence and the suggested grading (global explainability). They also gave a high helpfulness rating (6.00/7, Figure \ref{fig:rating}d) for the list of evidence provided by \xp. However, participants gave lower ratings on the probability (3.83/7, Figure \ref{fig:rating}f) and the confidence level (3.92/7, Figure \ref{fig:rating}g) elements in the mid-level samples because they were hard to read in \xp~--- 

\quo{``\textit{\dots these small words (pointing to the probability) \dots I didn't notice that very much \dots also it wasn't very easy to see.}''}{3}

The saliency map received a relatively higher rating (5.17/7, Figure \ref{fig:rating}h). However, some (P1, P5) participants found it hard to interpret the saliency map, especially for the cases where cues of attention were scattered across the entire evidence (see Figure \ref{fig:fail_evidence}a) ---

\quo{For the heatmap (the saliency map) \dots it is also a little bit confusing \dots it takes some time getting used to it and there are some false positives.}{1}

For the integrability dimension (\textbf{H3c}), participants gave overall higher scores for \xp~(Table \ref{tab:study_quant}, I1, \xp: $\mu$=5.91, ASAP: $\mu$=4.17, $p$=0.006). Specifically, participants were able to perform diagnoses based on the \xp's AI findings, which is similar to their workflow of collaborating with human trainees --- 

\quo{It's kind of like a first-year resident marking everything.}{1}

\quo{I'm a cytology fellow, and cases are pre-screened for us. And essentially this is doing similarly.}{4}

For the trust dimension, participants responded positively to \xp's capability of helping to grade meningiomas (T1: $\mu$=5.83) and their accuracy of the diagnoses while using the system (T2: $\mu$=6.00). However, some (P3, P4, P5) pointed out that they would spend more time examining the WSI entirely if more time had been granted --- 

\quo{I just went to the areas that the system suggested. If I had more time, I would like to just go to all the areas, just to feel more comfortable that I'm not missing anything.}{5}

Last, participants were more likely to use \xp~than the traditional interface (Table \ref{tab:study_quant}, F1, \xp: $\mu$=6.42, ASAP: $\mu$=3.75, $p$=0.002). Overall, 9/12 of the participants ``totally'' preferred \xp~over the traditional interface, while 3/12 ``much more'' preferred \xp~(Table \ref{tab:study_quant}, F2).

However, it is noteworthy that this study is based on participants' examination of WSIs, while pathologists use the optical microscope in their daily practice. During the study, 7/12 of our participants expressed that they preferred using an optical microscope with the glass slide \textit{vs.} a digital interface with the WSI --- \textit{``\dots it's much faster (in the microscope) than moving on the computer \dots we would prefer to look at a real slide instead of using a scan picture.''} (P2). As such, further comparison between \xp~and the optical microscope is considered future work.

\subsection{Recurring Themes}
\label{subsec:rec_the}
We analyzed the video recordings of the work sessions in a similar approach as described in Section \ref{subsec:existing}. Based on our observations of participants' using \xp~ and the interview with them, we discuss the following recurring themes that characterize how participants interacted with \xp. 

\subsubsection{Pathologists examine \xp's multiple criteria findings by prioritizing one and referring to others on demand}
We noted that participants tended to focus on a specific criterion. If that criterion alone did not meet the bar of a diagnosis for a higher grade, participants would use \xp~ to browse other criteria, looking for evidence of a differential diagnosis, until they identify sufficient evidence to support their hypothesis. 

\quo{I'm done. Because with the mitosis that high, you're done. You don't have to go through that stuff (other criteria).}{12}

However, some participants would also like to see other criteria and examine the slide comprehensively ---

\quo{With the mitosis rate that high, you don't actually need it (Ki-67) for the diagnosis. But I will have a look at it.}{1}

\quo{I will just look at (other criteria) because I don't want to grade by one single criterion (mitosis).}{3}

Such a relationship between criteria is analogous to `focus + context'  \cite{card1999readings} in information visualization --- different pathologists might focus on a few different criteria. Still, the other criteria are also important to serve as context at their disposal to support an existing diagnosis or find an alternative.

\subsubsection{\xp's top-down workflow with hierarchical explainable evidence enables pathologists to navigate between high-level AI results \& low-level WSI details}

One of the main reasons limiting the throughput of histopathological diagnosis is that criteria like mitotic count have very small size compared to the dimensions of WSIs. 
As a result, participants have to switch to high magnification to examine such small features in detail. Given the high resolution of the WSI, it is possible to `get lost' in the narrow scope of HPF, resulting in a time-consuming process to go through the entire WSI. With \xp, participants found its hierarchical design and the provision of mid-level evidence (\eg AI's ROI samples) the most helpful for diagnosis as it connects high-level findings and low-level details ---

\quo{It (\xp) finds the best area to look at. \dots You can jump there, and if it is a grade 3, then it is a grade 3. You don't have to look at other areas.}{6}

Furthermore, participants appreciated that \xp~provided heatmap visualizations to assist them in navigating the WSI out of the ROI samples ---

\quo{The heatmap is very useful to assist pathologists to go through the entire slide \dots which saves time and makes sure not missing anything.}{12}

\subsubsection{\xp's explainable design helps pathologists see what AI is doing} We found \xp's evidence-based justification of AI findings assisted participants in relating AI-computed results with evidence, which added explainability ---

\quo{System 2 (\xp) does find some evidence and assigns it to a particular observation that is related to the grading, so that it helps with explainability.}{3}

In \xp, the AI might make two types of mistakes that may incur potential bias: \one false positive, where AI mistakenly identifies negative areas as positive for a given criterion; \two false negative, where the AI misses positive areas corresponding to a criterion. We observed a number of false-positive detections that confused some participants. We also found out that the participants would rather deal with more false positives than false negatives so that signs of more severe grades would not be missed ---

\quo{It's better that it picks them up and gives me the opportunity to decline it.}{10}

Furthermore, although some participants found the saliency map hard to interpret in some cases, others used it to locate the cells that led to AI's grading --- 

\quo{There were a couple of instances where it was a bit more difficult to figure out what it (the saliency map) was trying to point out to me. But for the majority of the time, I could tell which area they (the saliency maps) were trying to show me.}{9}

Further, with the aid of the saliency map, participants could understand AI's limitations and what might have misled the AI ---

\quo{You can see what this system counted as mitosis \dots the heatmap (the saliency map) helps to understand why AI chose this or that area. For example, I think AI chose neutrophils as mitotic figures in some areas.}{6}

\subsubsection{Pathologists justify \xp~by incrementing human findings onto justified AI results} 

Given the explainable evidence provided by \xp, it was straightforward for participants to recognize and modify AI results when there was a disagreement. Specifically, participants could justify AI by clicking on the approve/decline/declare-uncertain buttons or modifying AI results directly on the criteria panel. If the justified AI results were sufficient to conclude a grading decision (\eg seven mitoses in 10 HPFs, enough to make the case as grade 2 (>4), but still far from grade 3 (<<20)), they would stop examining and report the grading. However, if the justified AI results appeared to be marginal (\eg 19 mitoses in 10 HPFs, which is only one mitosis away from upgrading the case to a grade 3), participants would continue to search based on the AI findings and add their new insights to grade ---

\quo{I count a total number of five \dots adding the previous 19 makes it 24 \dots this is grade 3.}{2}

What's more, for the cases where \xp~did not actively report positive detections, participants would examine the WSI manually as in a traditional interface --- that is, participants would use their experience to evaluate the case further and make a grading decision.

%% file: 10_discussion.tex
\section{Discussion}

In this section, we start by discussing this work's limitations and potential future improvements. We then summarize the design recommendations for future physician-AI collaborative systems. Finally, we focus on future directions for improving AI's integration into pathologists' workflow.

\subsection{Limitations \& Future Improvements}

We conclude the following limitations of our current work:
\begin{itemize}
    \item \xp~was evaluated on a small number of participants examining limited materials using a remote setup. As such, the observations and conclusions are inevitably biased and speculative;
    \item The AI's testing performance in this work was reported from an in-house dataset that was collected from one institute, while the evaluation of AI's alignment with the benchmarks from a large set of images from multiple medical centers was not conducted;
    \item \xp~currently does not support users to adjust the cut-off prediction threshold, hence resulting in an amount of false-positive evidence;
    \item Cases of the saliency map (see Figure \ref{fig:fail_evidence}) confuse some participants because they can not highlight cells appropriately;
\end{itemize}

Next, we will discuss the limitations and future improvements in detail.

\subsubsection{Increasing the scope of \xp's evaluation study} 

\begin{figure}
    \centering
    \includegraphics[width=0.7\linewidth]{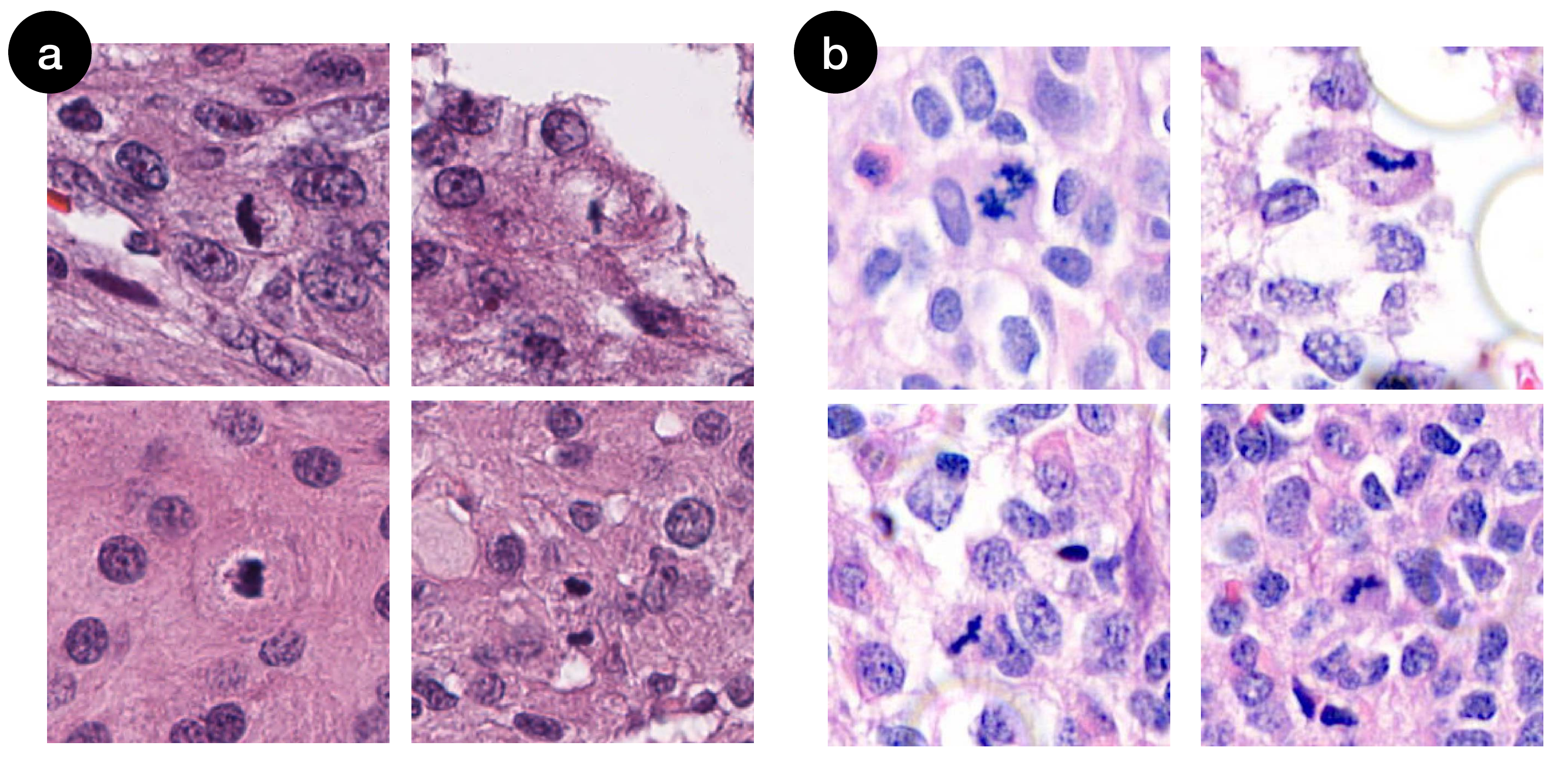}
    \caption{Mitoses from meningiomas (in x400), scanned by (a) the medical center in this study and (b) a different medical center. The difference in appearance is caused by the difference in processing procedures and scanners used.}
    \label{fig:medical_center}
\end{figure}

The scope of \xp's evaluation study was limited to the following four aspects:

\textbf{Study Material}. Due to the Institutional Review Board (IRB) regulations, only a limited number of images from one medical center were selected and used in \xp. This leaves the performance of \xp's AI questionable while being applied to images from other institutes. This is because other institutes might use a different staining process or a different type of scanner, causing a difference in the image domain/distribution (see Figure \ref{fig:medical_center}). Furthermore, the limited test cases generated for \xp's work sessions might not reflect the distributions of meningiomas in clinical settings.

\textbf{Participants: Recruitment and Sampling}. Because of the rare availability of medical professionals in neuropathology, we only recruited twelve participants for the study, most of whom were residents. This might cause the conclusions for \textbf{RQ1} and \textbf{RQ2} inevitably speculative because research has shown that pathologists' diagnostic accuracy might be related to their experience level \cite{ghezloo2022analysis}. Moreover, all participants came from one country, which might cause the qualitative observations to be biased since no pathologists from other countries were involved.

\textbf{Study Set-Up}. All studies were conducted online due to the COVID-19 pandemic. And the duration of each study (about 60 minutes) was relatively short in order to prove the long-term validity of \xp. Additionally, no clinical testing was conducted because of strict legislation regulations from US Food and Drug Administration (FDA).

\textbf{Apparatus}. The comparison between the \xp~and the optical microscope --- pathologists' first approach to seeing pathology slides, was not conducted. Although the FDA has lifted its restrictions on digital whole slide images for clinical use since 2017 \cite{fdawsi2017}, we found it is still challenging to persuade pathologists to move from the optical microscope to the digital interfaces (without AI): more than half of the participants expressed that they preferred using an optical microscope with the glass slide \textit{vs.} a traditional digital interface. Remarkably, participants found it challenging to navigate a digital whole slide image, which has also been described and discussed by Ruddle \etal~\cite{ruddle2016design}. However, our study found that pathologists preferred to use \xp~because it adds value to their workflow with AI. Therefore, we suggest that future medical systems highlight their benefit to pathologists as an incentive to overcome the limitations in traditional digital interfaces. 

In sum, future works should consider using more images from multiple medical centers, recruiting more participants with multiple experience levels, conducting long-term, in-person studies, and comparing \xp~with the optical microscope. With more data points collected, we can validate \xp's performance and generalizability more comprehensively.

\subsubsection{Enabling adjusting the thresholds within the interface}
\label{subsec:sensitivity}
Currently, \xp~ does not support directly changing the threshold for a positive result with the interface. 
In our user study, one participant mentioned that different pathologist might have different thresholds to call whether a piece of evidence is positive ---

\quo{``I only call the characteristic mitoses \dots other pathologists might have different thresholds.}{7}

Further, dealing with false positives and false negatives is another issue with the fixed-threshold scheme. 
From our study, we found out that pathologists would prefer high-sensitivity results that include some false positives rather than high-specificity results that have false negatives ---

\quo{I could have more faith if it could find all the candidates. And I could pretty easily click through and accept/reject, and know that it wasn't missing anything.}{8}

Therefore, the system, by default, should be designed to err on the side of caution, \eg showing a wide range of ROIs despite some being inevitably false positives. 
Pathologists are fast in examining ROIs (and ruling out false positives), whereas missing important features would come with a much higher cost (\eg delayed or missed treatment). 

\subsubsection{Improving the quality and granularity of explanations} 

\begin{figure}
    \centering
    \includegraphics[width=1.0\linewidth]{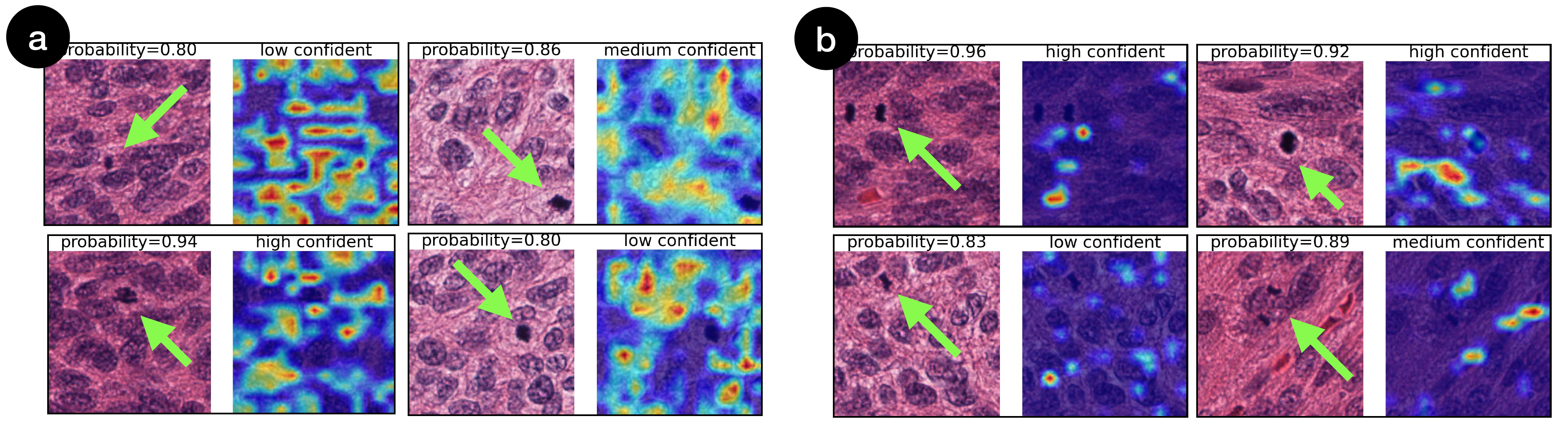}
    \caption{Examples of failure explanation cases, where the saliency shows (a) scattered attention across the image or (b) misleading hot spots. The green arrows point to the location of a mitosis figure marked by a human pathologist.}
    \label{fig:fail_evidence}
\end{figure}

In the study, we found a number of cases where the saliency maps failed to explain the classification predictions and caused confusion to the users. 
As shown in Figure \ref{fig:fail_evidence}, the failed saliency maps showed either scattered attention across the evidence (Figure \ref{fig:fail_evidence}a), or concentrated attention at the wrong place (Figure \ref{fig:fail_evidence}b). 
Such errors can be explained as the attention is reasoned from patch-wise annotations rather than localized ones because the localized annotations of positive findings are extremely labor-costly to obtain. 
The quality of the saliency maps can be potentially improved with the increment of training data for higher model generalization and the advent of the methodologies of unsupervised attention reasoning \cite{arrieta2020explainable}.  

Besides, knowing the location of a potential positive finding can be insufficient for pathologists. Since the pathological imaging of tissues is merely an approximation of the real condition, there can often exist uncertainty in diagnosis even for well-trained pathologists. As such, explaining why an area contains positive findings, \eg a highlighted cell is detected to stage as mitosis since its boundary is jagged, can be critical for systems in the future. Such causality enables a system to imitate how pathologists discuss with their peers, which can improve the collaboration between a system and its users. 
Moreover, future work should also employ more formal measurements (\eg System Causability Scale \cite{holzinger2020measuring}) to evaluate the quality of explanations.

\subsection{Design Recommendations for Physician-AI Collaborative Systems}
Although we focus on the grading of meningiomas in this work, we believe our two designs in \xp~--- joint-analyses of multiple criteria and explanation by hierarchically traceable evidence --- can be generalizable to other medical applications that require doctors to see and verify numerous criteria from various medical tests (such as grading astrocytoma, IDH mutant (WHO Grade 2-4),  solitary fibrous tumor (WHO Grade 1-3) \cite{10.1093/neuonc/noab106}). Here, we provide design recommendations for future physician-AI systems.

\subsubsection{Showing the logical relationships amongst multiple types of evidence at the top level} Carcinoma grading usually involves examining multiple criteria from various data sources (\eg H\&E slides, IHC test, FISH (fluorescence in situ hybridization) test, patient's health record). As such, one-size-fits-all AI models are not sufficient. In practice, multiple AI models are employed to locate different types of disease markers. To organize these AI-computed results, medical AI systems (such as \xp) should seek to present the logical relationship that connects these multiple criteria/features/sources of information and update final results dynamically given any pathologists' input (\eg acceptance or rejection of how AI computes each criterion). Such a design is more likely to match the clinical practice of pathologists and cost minimal extra learning when users onboard a system. 

\subsubsection{Making AI's findings traceable with hierarchically organized evidence} 
There is a pressing need to deal with the transparency of a black-box model and the traceability of the explanation evidence in high-stakes tasks (\eg medical diagnosis). As such, AI systems should provide local explainability where each piece of low-level evidence is traceable. In \xp, we employ the design of hierarchically traceable evidence for each criterion. Such an organization forms an `evidence chain' where each direct evidence is accountable for the high-level system output. Similar intuitions can also be applied to medical applications in a more general context, such as cancer staging \cite{lydiatt2017head} and cancer scoring \cite{humphrey2004gleason}, where the evidence is accumulated to arrive at a diagnosis.

\subsubsection{Employing a ``focus+context'' design toward presenting and/or interacting with multiple criteria} Medical diagnosis involves accumulating evidence from multiple criteria --- our study observed that pathologists started by focusing on one criterion while continuing to examine the others for a differential diagnosis. 
Thus, medical AI systems should make multiple criteria available, and support the navigation of such criteria following a ``focus+context'' design \cite{card1999readings}, which is commonly used in information visualization. The major design goal is to strike the dichotomy between juxtaposing the focused criterion with sufficient contextual criteria and overwhelming the pathologists with too much information. It is also possible for a system to, based on a patient's prior history and the pre-processing of their data, recommend a pathologist to start focusing on specific criteria followed by examining some others as context.

\subsection{On Integrating AI into Pathologists' Workflow}

\subsubsection{How has AI improved pathologists' diagnoses in \xp?}

Similar to previous human-AI collaborative research in medicine \cite{fogliato2022goes, lee2021human}, we discovered that using AI might improve pathologists' diagnosis quality. In pathology, the AI can  ``efficiently, systematically, exhaustively'' analyze the entire whole slide image \cite{steiner2020evaluation}. Therefore, \xp~ can help pathologist capture small-sized details they might miss in the manual examination, which can improve their sensitivity. \xp~further aggregates these details into AI-recommended regions of interest (ROIs), and pathologists can check each ROI of each criterion. Compared to the manual examinations where pathologists have to see multiple criteria with one pass (\ie ``multitasking'', as described in Section \ref{sec:formative}), such a design assists less-experienced pathologists in examining in a more organized, more comprehensive manner. 

Furthermore, \xp's ROI recommendations freed participants from heavy navigation and visual searching. Traditionally, pathologists navigate manually \cite{molin2015slide, ruddle2016design} and search visually to locate pathological patterns. With \xp, our participants could see and adjudicate ROI recommendations directly. However, it is noteworthy that forcing pathologists to see ROI recommendations might break their workflow. First, because ROI recommendations are not necessarily physically adjacent, pathologists need to ``jump'' from one ROI to another to examine them. And it is unclear whether pathologists can accept such `` ROI jumpings'' without continuous navigation (\ie panning and zooming). Second, the presentation of ROI recommendations (\eg in \xp, boxes) may also influence pathologists' judgement --- one participant expressed their concern when the ROI highlighted an area but failed to do so in a similar one --- ``\textit{If I called this positive (pointing at one recommendation box), should I also call this one (pointing at another area but not marked by recommendation boxes)?}''(P7). Hence, we suggest that future HCI systems study pathologists' acceptance of using ROIs to examine and elaborate more on the over-reliance issues.

\subsubsection{How to make human-AI systems in pathology more robust?} 
Although incorporating AI might benefit users, the performance of human-AI collaboration workflow might be influenced in clinical settings \cite{beede2020human, wang2021brilliant}. Therefore, it is crucial to design workflows that can cope with chaotic ``in the wild'' situations. \xp~applied two designs to assist pathologists to debug and refine the AI findings: \one hierarchical evidence that makes the AI analysis traceable and transparent; \two pathologists can refine the AI findings by approving/declining/declaring-uncertain AI analysis.

Based on the observations of how our participants interacted with \xp, we further discuss the potential approaches to make human-AI systems more robust for future pathology applications. The first approach is to add additional sources of information so that pathologists can verify the AI recommendations. For example, \xp~mimics how pathologists examine meningiomas and adds an additional test --- the Ki-67 test --- for mitosis ROI recommendations. In our user study, we found that pathologists could cross-check the Ki-67 hot-spot areas with mitosis ROI recommendations to validate the correctness.

For the systems without the luxury of additional tests, we suggest re-framing the human-AI collaboration workflow by forcing doctors to give a brief overview first and then retrieve AI recommendations on demand. Such a strategy is called the  ``\textit{cognitive forcing function}'' and is viable for reducing the over-reliance issues in previous literature \cite{buccinca2021trust}. We argue that such a workflow design is still integrable to pathologists' practice because their manual examination also starts with an overview of a slide \cite{ruddle2016design}. 

Finally, enabling users to control the recommendation process might also be a solution. For example, a slider can be used to control the sensitivities of AI-recommended ROIs. As such, pathologists can first see the most pressing ROI, and then gradually see more on demand. Such a design reduces the disruptive behavior of using AI systems in the wild and pathologists are more likely to accept it in practice \cite{cai2019human}.

\subsubsection{How should AI systems build trust for pathologists?} Previous HCI research advises informing doctors of AI's capabilities and limitations to gain trust \cite{cai2019hello}. For example, Sendak \etal~created a ``model fact sheet'' inspired by pharmaceutical drug labels to inform doctors of AI details \cite{sendak2020human}. In our study, we also discovered that participants preferred to know the AI capabilities --- ``\textit{I really wanna cross-check (AI's) accuracy with a human observer, and cases of a range of mitosis, from rare mitosis to frequent mitosis.}''(FP1) \textit{``Pathologists are data-driven ... if you can show it (AI) is accurate for like 1,000 cases, they may buy it.''}(P1) As such, we suggest future medical AI systems to demonstrate AI's capabilities by presenting with a set of examples with AI's predictions and ground truth. With the help of examples, pathologists can briefly evaluate AI performance and know its capabilities and limitations.

Apart from AI's information, previous studies indicate that explanations might improve trust: some attempt to explain AI predictions with XAI components (\eg the saliency map \cite{zhang2019pathologist}), while others build inherently interpretable models (\eg concept bottleneck models \cite{koh2020concept}). During the study, we found that our participants preferred simple explanations during the interaction with \xp. Although complex explanations (\eg concept explanations) might provide a more detailed background, pathologists might justify a vast number of explanations during the time-pressing diagnosis process. If the explanations cannot capture pathologists' attention initially, they might ignore them for the rest of the examination process (also described by P3 in our user study). Therefore, we suggest future medical AI systems allow pathologists to see \textit{levels of} explanations on demand. For example, pathologists might see simple visual explanations by default but can opt to see more detailed explanations if they wish.

%% file: 11_conclusion.tex
\section{Conclusion}
In this work, we identify three challenges of comprehensiveness, explainability, and integrability that prevent AI from being adopted in a complex clinical setting for pathologists. To close these gaps, we implement \xp~with two key design ingredients: \one joint-analyses of multiple criteria and \two explanation by hierarchically traceable evidence. To validate \xp, we conducted work sessions with twelve medical professionals in pathology across three medical centers. Our findings suggest that \xp~can leverage AI to reduce pathologists' cognitive workload for meningioma grading. Meanwhile, pathologists benefited from the design and made fewer mistakes with \xp, compared to the manual baseline interface. By observing pathologists' use of \xp~and collecting their quantitative and qualitative feedback, we indicate how pathologists may collaborate with AI and summarize design recommendations. We believe that \xp~ is useful for other HCI research by providing first-hand information on how pathologists collaborate and manage multiple AI outcomes, which opens up a new space for pathologist-AI interaction possibilities.

%% file: 13_appendix.tex
\section{WHO Guidelines for Meningioma Grading}
\label{sec:mg-grade}

As specified by the World Health Organization (WHO) guidelines \cite{louis20072007, louis20212021}, meningioma grading diagnosis can be based on the following criteria:

\begin{itemize}
    \item \textbf{Grade 1} (benign) meningiomas include ``histological variant other than clear cell, chordoid, papillary, and rhabdoid ''\cite{brat2008surgical} with some exceptions \textit{and} a lack of criteria for grade 2 and 3 meningiomas.
    \item \textbf{Grade 2} (formerly called atypical) meningiomas are recognized by meeting at least one of the four following criteria:
    \begin{enumerate}[wide=\dimexpr\parindent+\labelsep\relax, leftmargin=* ]
        \item 
        The presence of  $\geq$ 2.5 mitoses/mm$^2$ (equating to $\geq$ 4 mitoses per/10 high power field (HPF) of 0.16 mm$^2$.
        Moreover, since mitoses are challenging to recognize in H\&E, the Ki-67-positive nuclei (Figure \ref{fig:meningioma}k) in the corresponding areas of Ki-67 (Figure \ref{fig:meningioma}d) are often compared for disambiguation;
        \item At least three out of five following histopathological features are observed: hypercellularity --- an abnormal excess of cells in the specimen (Figure \ref{fig:meningioma}f), prominent nucleoli --- enlarged nucleoli in a cell (usually as a cluster) (Figure \ref{fig:meningioma}g,m), sheeting --- loss of `whirling' architecture (Figure \ref{fig:meningioma}h), necrosis --- irreversible injury to cells (Figure \ref{fig:meningioma}i), and small cell --- cluster of cells with high nuclear/cytoplasmic ratio (Figure \ref{fig:meningioma}j);
        \item Brain invasion --- invasive tumor cells within the brain tissue is observed (Figure \ref{fig:meningioma}e);
        \item The dominant appearance of clear cell or chordoid subtype.
    \end{enumerate}
    \item \textbf{Grade 3} meningiomas are decided if at least one of the following criteria met \cite{backer2012histopathological, louis20212021}:
    \begin{enumerate}[wide=\dimexpr\parindent+\labelsep\relax, leftmargin=* ]
        \item Mitotic figures of $\geq$ 12.5 mitoses/mm $^2$ (equal to $\geq$ 20 mitoses/10 HPF of 0.16 mm$^2$);
        \item The appearance of frank anaplasia, papillary or rhabdoid subtype with some exceptions;
        \item Molecular alterations, such as a \textit{TERT} promoter mutation; and/or homozygous \textit{CDKN2A} and/or \textit{CDKN2B} deletion.
    \end{enumerate}
\end{itemize}